\documentclass[12pt]{article}
\usepackage{amssymb}

\usepackage{epsfig}
\usepackage{graphics}
\usepackage{amsmath}

\begin{document}

\title{{\Large Dirac fermions in a magnetic-solenoid field}}
\author{S. P. Gavrilov\thanks{%
Dept. Fisica e Quimica, UNESP, Campus de Guaratingueta, Brazil; On leave
from Tomsk State Pedagogical University, 634041 Russia; e-mail:
gavrilovsp@hotmail.com}, D. M. Gitman\thanks{%
e-mail: gitman@dfn.if.usp.br}, A. A. Smirnov\thanks{%
e-mail: saa@dfn.if.usp.br}, and B. L. Voronov\thanks{%
Lebedev Physical Institute, Moscow, Russia; e-mail: voronov@lpi.ru }}
\date{\today }
\maketitle

\begin{abstract}
We consider the Dirac equation with a magnetic-solenoid field (the
superposition of the Aharonov--Bohm solenoid field and a collinear uniform
magnetic field). Using von Neumann's theory of the self-adjoint extensions
of symmetric operators, we construct a one-parameter family and a
two-parameter family of self-adjoint Dirac Hamiltonians in the respective $%
2+1$ and $3+1$ dimensions. Each Hamiltonian is specified by certain
asymptotic boundary conditions at the solenoid. We find the spectrum and
eigenfunctions for all values of the extension parameters. We also consider
the case of a regularized magnetic-solenoid field (with a finite-radius
solenoid field component) and study the dependence of the eigenfunctions on
the behavior of the magnetic field inside the solenoid. The zero-radius
limit yields a concrete self-adjoint Hamiltonian for the case of the
magnetic-solenoid field. In addition, we consider the spinless particle in
the regularized magnetic-solenoid field. By the example of the radial Dirac
Hamiltonian with the magnetic-solenoid field, we present an alternative,
more simple and efficient, method for constructing self-adjoint extensions
applicable to a wide class of singular differential operators.
\end{abstract}

\section{Introduction}

In the present Chapter, we study the Dirac particle in a magnetic-solenoid
field. This field is the superposition of the Aharonov--Bohm (AB) field (the
field of an infinitely long and infinitesimally thin solenoid) and a
collinear uniform magnetic field. We note that the solutions of the Schr\"{o}%
dinger, Klein-Gordon, and Dirac equations in such a field were already
discussed in \cite{L83,BGS86,SS89,BGT01}. In particular, the solutions of
the Dirac equation with the magnetic-solenoid field in $2+1$ and $3+1$
dimensions were studied in detail in \cite{BGT01}. These solutions were used
to calculate various characteristics of the particle radiation in such a
field \cite{BGLT01}. In fact, the AB effect in the synchrotron radiation was
investigated. However, a number of important and interesting aspects related
to the rigorous quantum-mechanical treatment of the solutions of the Dirac
equation with the singular magnetic-solenoid field were not considered. In
particular, it was pointed out that a critical subspace exists where the
problem of self-adjointness for the Dirac Hamiltonian taken naively arises 
\cite{BGT01}. This problem and the associated problem of the completeness of
the solutions was not solved.

We note that even for the pure AB field it was not simple to solve these two
problems inherent in both the relativistic and nonrelativistic cases. The
construction of self-adjoint nonrelativistic Hamiltonians for the AB-field
case by the method of the self-adjoint extensions of symmetric operators was
first studied in detail in \cite{T74} where the regularized AB field was
also considered. The self-adjoint extension method was also invoked in the
anyon physics \cite{MT91,GHK91,GMS96}. The self-adjointness problem and the
need for self-adjoint extension method in the case of the Dirac Hamiltonian
with the pure AB field in $2+1$ dimensions was recognized in \cite{GJ89,G89}%
. The interaction between the magnetic momentum of a charged particle and
the AB field essentially changes the behavior of the wave functions at the
magnetic string \cite{G89,H90,H91}. It was shown that there exists a
one-parameter family of self-adjoint extensions specified by certain
boundary conditions at the origin; in what follows, we call such boundary
conditions the self-adjoint boundary conditions. Self-adjoint extensions for
the case of the Dirac Hamiltonian in $3+1$ dimensions were constructed in 
\cite{VGS91}. An alternative method for solving the Hamiltonian extension
problem in $2+1$ and in $3+1$ dimensions was presented in \cite{AJS95,AJS96}%
. It was shown that in $2+1$ dimensions, only two values of the extension
parameter correspond to the presence of the point-like magnetic field at the
origin, whereas other values of the extension parameter correspond to
additional contact interactions \cite{MT93}. One possible self-adjoint
boundary condition was obtained in \cite{H90,AMW89,FL91} by specifically
regularizing the Dirac delta function, requiring the continuity of the both
Dirac spinor components at a finite radius, and then shrinking this radius
to zero. Other extensions in $2+1$ and $3+1$ dimensions were constructed in 
\cite{BFS99,BFKS00,FK01} by imposing spectral boundary conditions of the
Atiyah-Patodi-Singer type \cite{APS75} (the MIT boundary conditions) at a
finite radius, and then taking the zero-radius limit. It was shown that for
some extension parameters it is possible to find a domain where the
Hamiltonian is self-adjoint and commutes with the helicity operator \cite
{CP94,ACP01}. The bound state problem for particles with magnetic moment in
the AB potential was considered in detail in \cite{BV93,VB94,BV94}. The
physically motivated boundary conditions for the particle scattering by the
AB field and a Coulomb center were studied in \cite{CP93,H93,HP96}.

It is unclear whether the  self-adjointness problem for the case of the
magnetic-solenoid field can be automatically solved by directly extending
the results for the pure AB field, i.e., by applying the same boundary
conditions because the presence of the uniform magnetic field essentially
changes the domains of the relevant operators, in particular, the deficiency
subspaces, and changes the energy spectrum of the spinning particle from
continuous to discrete. Therefore,\ the case of the magnetic-solenoid field
requires an independent study. By analogy with the pure AB field, it also
seems important to consider the regularized magnetic-solenoid field (we call
the regularized magnetic-solenoid field the superposition of a uniform
magnetic field and the regularized AB field) and to study the solutions of
the Dirac equation in such a field. This problem was not solved before and
is of particular interest irrespective of the extension problem. The Pauli
equation with the magnetic-solenoid field was recently studied in \cite
{T00,C00}. The problem of defining the Hamiltonian in a particular case of
the magnetic-solenoid field (both fields have the same direction) was
considered in \cite{ESV02} (the scalar case) and in \cite{FP01} (the
spinning case in $2+1$ dimensions). We note that the AB symmetry is violated
for the spinning particle case, which is therefore sensible to the solenoid
flux direction. As a consequence some of the results can depend on the
mutual orientation, parallel or antiparallel, of the solenoid flux and the
uniform magnetic field: either parallel or antiparallel. We study both
possibilities in detail. The $3+1$ dimensional spinning problem and the
relation of the self-adjoint extensions to the regularized problems were not
studied.

In this Chapter we consider the Dirac equation with the general
magnetic-solenoid field (the uniform magnetic field and the AB field can
have both the same and opposite directions) and with the regularized
magnetic-solenoid field in $2+1$ and $3+1$ dimensions. We start with
separating the variables and describing the exact square-integrable
solutions of the radial Dirac equation in the magnetic-solenoid field in $2+1
$ dimensions. Using von Neumann's theory of the \thinspace\ self-adjoint
extensions of symmetric operators, we then construct a one-parameter family
of \thinspace\ self-adjoint Dirac Hamiltonians specified by boundary
conditions at the AB solenoid and find the spectrum and eigenfunctions for
each value of the extension parameter. We reduce the $\left( 3+1\right) $%
-dimensional problem to the $\left( 2+1\right) $-dimensional one by a proper
choice of the spin operator, which allows realizing all the programme of
constructing  self-adjoint extensions and finding spectra and eigenfunctions
in the previous terms. We then turn to the regularized case of finite-radius
solenoid. We study the structure of the corresponding eigenfunctions and
their dependence on the behavior of the magnetic field inside the solenoid.
Considering the zero-radius limit with the fixed value of the magnetic flux,
we obtain a concrete  self-adjoint Hamiltonian corresponding to a specific
boundary condition for the case of the magnetic-solenoid field with the AB
solenoid. For completeness we also study the behavior of the spinless
particle in the regularized magnetic-solenoid field.

It is well worth noting that in our particular case, as in many other cases
where singular differential operators occur, the general von Neumann's
procedure for the  self-adjoint extensions of symmetric operators can be
significantly reduced to analyzing the behavior of the corresponding
functions in the vicinity of the singularity. In Sec. 5, we present this
more simple method for the  self-adjoint extensions as applied to the radial
Dirac Hamiltonian with the magnetic-solenoid field.

\section{Exact solutions}

We consider the Dirac equation ($c=\hbar =1$) in $3+1$ and $2+1$ dimensions, 
\begin{equation}
i\partial _{0}\Psi =H\Psi ,\;H=\gamma ^{0}\left( \mbox{\boldmath$\gamma$%
\unboldmath}\mathbf{P}+M\right) \,.  \label{abe1}
\end{equation}
Here, $\gamma ^{\nu }=\left( \gamma ^{0},\mbox{\boldmath$\gamma$\unboldmath}%
\right) ,\;\mbox{\boldmath$\gamma$\unboldmath}=\left( \gamma ^{k}\right) $,\ 
$P_{k}=i\partial _{k}-qA_{k},$ $k=1,2$ for $2+1$ and $k=1,2,3$ for $3+1$,$%
\;\nu =\left( 0,k\right) ;\;q$ is an algebraic charge, for electrons, $%
q=-e<0 $. As an external electromagnetic field, we take the
magnetic-solenoid field. The magnetic-solenoid field is the collinear
superposition of the constant uniform magnetic field $B$ and the
Aharonov-Bohm field $B^{AB}$ (the AB field is a field of an infinitely long
and infinitesimally thin solenoid). The complete Maxwell tensor is 
\begin{equation*}
F_{\lambda \nu }=\overline{B}\left( \delta _{\lambda }^{2}\delta _{\nu
}^{1}-\delta _{\lambda }^{1}\delta _{\nu }^{2}\right) ,\;\overline{B}%
=B^{AB}+B\,.
\end{equation*}
The AB\ field is singular at $r=0$, 
\begin{equation*}
B^{AB}=\Phi \delta (x^{1})\delta (x^{2})\,.
\end{equation*}
The AB\ field creates the magnetic flux $\Phi $. It is convenient to
represent this flux as: 
\begin{equation}
\Phi =\left( l_{0}+\mu \right) \Phi _{0},\Phi _{0}=2\pi /e\;,  \label{abex1}
\end{equation}
where $l_{0}$ is integer and $0\leq \mu <1$.

If we use the cylindric coordinates $\varphi $ and $r$:$\;x^{1}=r\cos
\varphi $, $x^{2}=r\sin \varphi $, then the potentials have the form 
\begin{eqnarray}
&&A_{0}=0,\;eA_{1}=\left[ l_{0}+\mu +A\left( r\right) \right] \frac{\sin
\varphi }{r},\;eA_{2}=-\left[ l_{0}+\mu +A\left( r\right) \right] \frac{\cos
\varphi }{r}\;,  \notag \\
&&\,(A_{3}=0\;\mathrm{in\;}3+1{}),\;\;A\left( r\right) =eBr^{2}/2\;.
\label{abe2}
\end{eqnarray}

We conventionally treat the Dirac equation quantum-mechanically as the
evolution Schr\"{o}dinger-type equation with the Hamiltonian $H$, an
operator in the appropriate Hilbert space, and restrict ourselves to the
stationary solutions, i.e., to the (generalized) eigenfunctions of the Dirac
Hamiltonian. This implies that we seek the stationary solutions that are
bounded at infinity and locally square integrable.

\subsection{Solutions in 2+1 dimensions}

We first consider the problem in $2+1$ dimensions There are two
non-equivalent representations for the $\gamma $-matrices in $2+1$
dimensions: 
\begin{equation*}
\gamma ^{0}=\sigma ^{3},\;\gamma ^{1}=i\sigma ^{2},\;\gamma ^{2}=-i\sigma
^{1}\zeta ,\;\;\zeta =\pm 1\,,
\end{equation*}
where the ''polarizations'' $\zeta =\pm 1$ correspond to the respective
''spin up'' and ''spin down'' particles and $\mbox{\boldmath$\sigma$%
\unboldmath}=\left( \sigma ^{i}\right) $ are the Pauli matrices. The
stationary case is assigned the following form of the spinors $\Psi (x):$%
\begin{equation}
\Psi (x)=\exp \left\{ -i\varepsilon x^{0}\right\} \psi _{\varepsilon
}^{(\zeta )}\left( x_{\perp }\right) \,,\;\zeta =\pm 1,\;x_{\perp }=\left(
0,x^{1},x^{2}\right) \,.  \label{abe6}
\end{equation}
The stationary Dirac equation in the both representations is: 
\begin{eqnarray}
&&\left( \mbox{\boldmath$\sigma$\unboldmath}\mathbf{P}_{\perp }+M\sigma
^{3}\right) \psi _{\varepsilon }^{(1)}(x_{\perp })=\varepsilon \psi
_{\varepsilon }^{(1)}(x_{\perp }),\;P_{\perp }=\left( 0,P_{1},P_{2}\right)
\,,  \label{abe7a} \\
&&\left( \sigma ^{1}\mbox{\boldmath$\sigma$\unboldmath}\mathbf{P}_{\perp
}\sigma ^{1}+M\sigma ^{3}\right) \psi _{\varepsilon }^{(-1)}(x_{\perp
})=\varepsilon \psi _{\varepsilon }^{(-1)}(x_{\perp })\,.  \label{abe7b}
\end{eqnarray}
We note that the energy eigenvalues can be positive, $\varepsilon
={}_{+}\varepsilon >0$ , or negative, $\varepsilon ={}_{-}\varepsilon <0$.
We also can see that (\ref{abe7a}) and (\ref{abe7b}) are related by 
\begin{equation}
\psi _{\varepsilon }^{(-1)}(x_{\perp })=\sigma ^{2}\psi _{-\varepsilon
}^{(1)}(x_{\perp })\,.  \label{abe8}
\end{equation}
In what follows, we use the representation defined by $\zeta =1$.

As the total angular momentum operator, we choose $J=-i\partial _{\varphi
}+\sigma ^{3}/2$ that is the dimensional reduction of the operator $J^{3}$
in $3+1$ dimensions. The operator $J$ commutes with the Hamiltonian $H$.
Therefore, we can consider Eq. (\ref{abe7a}) separately in each of
eigenspaces of the operator $J,$%
\begin{equation}
J\psi _{\varepsilon }^{(1)}\left( x_{\perp }\right) =\left( l-l_{0}-\frac{1}{%
2}\right) \psi _{\varepsilon }^{(1)}\left( x_{\perp }\right) \,,\;l\in 
\mathbb{Z\,}.  \label{abe10}
\end{equation}
Representing the spinors $\psi _{\varepsilon }^{(1)}$ in the form 
\begin{equation}
\psi _{\varepsilon }^{(1)}\left( x_{\perp }\right) =g_{l}(\varphi )\psi
_{l}\left( r\right) ,\;g_{l}(\varphi )=\frac{1}{\sqrt{2\pi }}\exp \left\{
i\varphi \left[ l-l_{0}-\frac{1}{2}\left( 1+\sigma ^{3}\right) \right]
\right\} \,,  \label{abe11}
\end{equation}
we reduce equation \ref{abe7a} to the radial Dirac equation for the radial
spinor $\psi _{l}\left( r\right) $, 
\begin{eqnarray}
&&h\psi _{l}(r)=\varepsilon \psi _{l}(r)\,,\;\;h=\Pi +\sigma ^{3}M\,,
\label{abe12} \\
&&\Pi =-i\left\{ \partial _{r}+\frac{\sigma ^{3}}{r}\left[ \mu +l-\frac{1}{2}%
\left( 1-\sigma ^{3}\right) +A\left( r\right) \right] \right\} \sigma ^{1}\;,
\label{abe13}
\end{eqnarray}
with the radial Hamiltonian\footnote{%
By the radial Hamiltonian $h,$ we mean the whole family of $h$ with
different $l=0,\pm 1,...$ and arbitrary $B.$} $h$; $\Pi $ defines the action
of the spin projection operator on the radial spinor in the subspace with a
given $l$, 
\begin{equation*}
\mbox{\boldmath$\sigma$\unboldmath}\mathbf{P}_{\perp }g_{l}(\varphi )\psi
_{l}\left( r\right) =g_{l}(\varphi )\Pi \psi _{l}\left( r\right) \,.
\end{equation*}

It is convenient to represent the radial spinor as 
\begin{equation}
\psi _{l}(r)=\left[ \sigma ^{3}\left( \varepsilon -\Pi \right) +M\right]
u_{l}(r)\,,  \label{abe15}
\end{equation}
where 
\begin{eqnarray}
&&u_{l}(r)=\sum_{\sigma =\pm 1}c_{\sigma }u_{l,\sigma }(r)\,,\;u_{l,\sigma
}(r)=\phi _{l,\sigma }(r)\upsilon _{\sigma }\,,  \notag \\
&&\upsilon _{1}=\left( 
\begin{array}{l}
1 \\ 
0
\end{array}
\right) ,\;\;\upsilon _{-1}=\left( 
\begin{array}{l}
0 \\ 
1
\end{array}
\right) ,  \label{abe16}
\end{eqnarray}
and $c_{\sigma }$ are some constants. It follows from (\ref{abe12}) that $%
\Pi ^{2}u=\left( \varepsilon ^{2}-M^{2}\right) u$; therefore, the radial
functions $\phi _{l,\sigma }(r)$ satisfy the equation 
\begin{eqnarray}
&&\left\{ \rho \frac{d^{2}}{d\rho ^{2}}+\frac{d}{d\rho }-\frac{\rho }{4}+%
\frac{1}{2}\left[ \frac{\omega }{\gamma }-\xi \left( \mu +l-\frac{1}{2}%
\left( 1-\sigma \right) \right) \right] -\frac{\nu ^{2}}{4\rho }\right\}
\phi _{l,\sigma }(r)=0\,,  \label{abe17} \\
&&\rho =\gamma r^{2}/2,\;\;\gamma =e\left| B\right| ,\;\;\xi =\mathrm{sgn}%
B,\;\;\nu =\mu +l-\left( 1+\sigma \right) /2,\;\omega =\varepsilon
^{2}-M^{2}\,.  \notag
\end{eqnarray}
Solutions of the equation (\ref{abe17}) were studied in \cite{BGT01}. The
results can be summarized as follows.

For any $l$, there exists a set of solutions $\phi _{l,\sigma }=(\phi
_{m;l,\sigma }\,$, $\;m=0,1,2,\ldots )$, 
\begin{equation}
\phi _{m,l,\sigma }(r)=I_{m+|\nu |,m}\left( \rho \right) \,,  \label{abe19}
\end{equation}
square integrable and regular\footnote{%
Here, we use the terms ''regular'' and ''irregular'' at $r=0$ in the
following sense. We call a function to be regular if it behaves as $r^{c}$
at $r=0$ with $c\geq 0$, and irregular if $c<0$. We call a spinor to be
regular when all its components are regular, and irregular when at least one
of its components is irregular.} at $r=0$. Here $I_{n,m}(\rho )$ are the
Laguerre functions that are presented in Appendix A.

For $l=0$, there exist solutions square integrable and irregular at $r=0$ if 
$\mu \neq 0$. A general irregular solution for $l=0$ and $\mu \neq 0$ is 
\begin{eqnarray}
&&\phi _{\omega ,\sigma }(r)=\psi _{\lambda ,\alpha }(\rho )=\rho
^{-1/2}W_{\lambda ,\alpha /2}(\rho )\,,  \notag \\
&&\alpha =\mu -\left( 1+\sigma \right) /2,\;2\lambda =\omega /\gamma -\xi 
\left[ \mu -\left( 1-\sigma \right) /2\right] \,,  \label{abe25}
\end{eqnarray}
where $W_{\lambda ,\alpha /2}$ are the Whittaker functions (see \cite{GR94}
, 9.220.4). The spinors in (\ref{abe12}) that are constructed using these
functions\ are square integrable for arbitrary complex $\lambda $. The
functions $\psi _{\lambda ,\alpha }$ were studied in detail in \cite{BGT01},
some important relations for these functions are presented in Appendix A. We
see that interpretation of $\omega $\ as energy is impossible for complex $%
\lambda $. For real $\lambda $\ there exist a set of solutions (\ref{abe25})
which can be expressed in terms of the Laguerre functions with integer
indices: 
\begin{eqnarray}
&&\phi _{m,1}^{ir}(r)=I_{m+\mu -1,m}\left( \rho \right) \,,\;\sigma
=1{},\;m=0,1,2,\ldots ,  \notag \\
&&\phi _{m,-1}^{ir}(r)=I_{m-\mu ,m}\left( \rho \right) \,,\;\sigma
=-1,\;m=0,1,2,\ldots \,.  \label{abe20}
\end{eqnarray}
All the corresponding solutions $\psi _{l}(r)$ of Eq. (\ref{abe12}) are
square integrable on the half-line with the measure $rdr$. The Laguerre
functions in Eqs. (\ref{abe19}), (\ref{abe20}) are expressed via the
Laguerre polynomials.

Eigenvalues $\omega $ and the form of the spinors depend on $\mathrm{sgn}B$.
In what follows, we present the results for $B>0$. The results for $B<0$
cannot be obtained trivially from the ones for $B>0$. We present them in
Appendix B. The spectrum of $\omega $ corresponding to the functions $\phi
_{m,l,\sigma }(r)$ is 
\begin{equation}
\omega =\left\{ 
\begin{array}{l}
2\gamma \left( m+l+\mu \right) ,\;{}l-\left( 1+\sigma \right) /2\geq 0 \\ 
2\gamma \left( m+\left( 1+\sigma \right) /2\right) ,\;{}l-\left( 1+\sigma
\right) /2<0
\end{array}
\right. ,  \label{abe21}
\end{equation}
and the spectrum of $\omega $ corresponding to the functions $\phi
_{m,\sigma }^{ir}(r)$ is 
\begin{equation}
\omega =\left\{ 
\begin{array}{l}
2\gamma \left( m+\mu \right) ,\;\sigma =1 \\ 
2\gamma m,\;\sigma =-1
\end{array}
\right. .  \label{abe22}
\end{equation}

We require that the spinors $u_{l}\left( r\right) $ be eigenvectors for $\Pi 
$, such that the functions $u_{m,l,\pm }$ satisfy the equation 
\begin{equation}
\Pi u_{m,l,\pm }(r)=\pm \sqrt{\omega }u_{m,l,\pm }(r)\,.  \label{abe26}
\end{equation}
We now can specify the coefficients in (\ref{abe16}).

In the case $\omega =0$, we have 
\begin{equation}
u_{0,l}(r)=\left( 
\begin{array}{l}
0 \\ 
\phi _{0,l,-1}(r)
\end{array}
\right) ,\;{}l\leq -1;\;\;\;u_{0}^{I}(r)=\left( 
\begin{array}{l}
0 \\ 
\phi _{0,-1}^{ir}(r)
\end{array}
\right) ,\;{}l=0\;.  \label{abe27}
\end{equation}
This can be easily seen from the relations (\ref{4.11}) - (\ref{4.16}) for
the Laguerre functions $I_{n,m}(\rho )$ .

In the case $\omega \neq 0$, we have 
\begin{eqnarray}
&&u_{m,l,\pm }(r)=\left( 
\begin{array}{l}
\phi _{m,l,1}(r) \\ 
\pm i\phi _{m,l,-1}(r)
\end{array}
\right) \,,\mathrm{\;}l\geq 1,\;\omega =2\gamma \left( m+l+\mu \right) \,, 
\notag \\
&&u_{m+1,l,\pm }(r)=\left( 
\begin{array}{l}
\phi _{m,l,1}(r) \\ 
\mp i\phi _{m+1,l,-1}(r)
\end{array}
\right) \,,\mathrm{\;}l\leq -1,\;\omega =2\gamma \left( m+1\right) \,, 
\notag \\
&&u_{m+1,\pm }^{I}(r)=\left( 
\begin{array}{l}
\phi _{m,0,1}(r) \\ 
\mp i\phi _{m+1,-1}^{ir}(r)
\end{array}
\right) ,\mathrm{\;}l=0,\;\omega =2\gamma \left( m+1\right) \,,  \notag \\
&&u_{m,\pm }^{II}(r)=\left( 
\begin{array}{l}
\phi _{m,1}^{ir}(r) \\ 
\pm i\phi _{m,0,-1}(r)
\end{array}
\right) ,\;l=0,\;\omega =2\gamma \left( m+\mu \right) \,.  \label{abe28}
\end{eqnarray}

For $\omega \neq 0$, we construct solutions of the Dirac equation using the
spinors $u$ corresponding to the positive eigenvalues of the operator $\Pi $%
. These solutions are 
\begin{eqnarray}
&&\psi _{m,l}(r)=N\left[ \sigma ^{3}\left( \varepsilon -\sqrt{\omega }%
\right) +M\right] u_{m,l,+}(r),\;l\neq 0\,,  \notag \\
&&\psi _{m}^{I,II}(r)=N\left[ \sigma ^{3}\left( \varepsilon -\sqrt{\omega }%
\right) +M\right] u_{m,+}^{I,II}(r),\;l=0\,,  \label{abe31}
\end{eqnarray}
where $N$ is a normalization constant. Substituting (\ref{abe31}) into (\ref
{abe12}), we obtain the two types of states corresponding to particles, $%
\,_{+}\psi $, and antiparticles, $\,_{-}\psi $, with $\varepsilon ={}_{\pm
}\varepsilon =\pm \sqrt{M^{2}+\omega }$, respectively. The particle and
antiparticle spectra are symmetric, that is, $|_{+}\varepsilon
|=|_{-}\varepsilon |$, for the given quantum numbers $m$ and $l$.

We now consider the case $\omega =0$. As it follows from (\ref{abe12}) and (%
\ref{abe27}), only the negative energy solutions (antiparticles) are
possible. They coincide with the corresponding spinors $u$ up to a
normalization constant, 
\begin{equation}
_{-}\psi _{0,l}(r)=Nu_{0,l}(r){},\;l\leq -1;\;\;\;_{-}\psi
_{0}^{I}(r)=Nu_{0}^{I}(r){},\;l=0\,.  \label{abe32}
\end{equation}
Thus, only antiparticles have the rest energy level. The particle\ lower
energy level for $l\leq 0$ is $_{+}\varepsilon =\sqrt{M^{2}+2\gamma }$.

All the radial spinors $\psi _{m,l}(r)$ are orthogonal for different $m$.
The same is true for both the spinors $\psi _{m}^{I}$ and $\psi _{m}^{II}$.
In the general case, the spinors of the different types are not orthogonal.
Using Eq. (\ref{4.41}) of Appendix A, we can prove this fact and also
calculate the normalization factor which has the same form for all types of
the spinors, 
\begin{equation}
N=\sqrt{\frac{\gamma }{2\left[ \left( \varepsilon -\sqrt{\omega }\right)
^{2}+M^{2}\right] }}\,.  \label{abe33b}
\end{equation}

In addition, on the subspace $l=0,$ there are solutions of Eq. (\ref{abe12})
that are expressed via the functions $\psi _{\lambda ,\alpha }(\rho )$ (\ref
{abe25}). We represent these solutions as 
\begin{eqnarray}
&&\psi _{\omega }(r)=\left[ \sigma ^{3}\left( \varepsilon -\Pi \right) +M%
\right] u_{\omega }(r)\,,  \notag \\
&&u_{\omega }(r)=c_{1}u_{\omega ,1}(r)+c_{-1}u_{\omega ,-1}(r),\;u_{\omega
,\sigma }(r)=\phi _{\omega ,\sigma }(r)v_{\sigma }\,.  \label{abe34}
\end{eqnarray}
Using the relations (\ref{4.145}) for the functions $\psi _{\lambda ,\alpha
}(\rho )$, we obtain the useful expressions 
\begin{equation}
\Pi u_{\omega ,1}(r)=i\sqrt{2\gamma }u_{\omega ,-1}(r),\;\Pi u_{\omega
,-1}(r)=-i\frac{\omega }{\sqrt{2\gamma }}u_{\omega ,1}(r)\,,  \label{abe35}
\end{equation}
Using Eq. (\ref{4.148}) from Appendix A, we can see that the spinors $\psi
_{\omega }(r)$ and $\psi _{\omega ^{\prime }}(r)$, $\omega \neq \omega
^{\prime },$ are not orthogonal in the general case.

The completeness of all the obtained solutions is also an open problem.

We conventionally treat the obtained solutions and spectrum as the
respective energy eigenvectors and energy eigenvalues for the Dirac
Hamiltonian $H$. In view of the confining uniform magnetic field and the
apparent self-adjointness of $H$, we expect bound states and real discrete
spectrum. What concerns the subspaces $l\neq 0,$ our expectations are
completely realized. But in the subspace $l=0$ for $\mu \neq 0$, there exist
square-integrable solutions with complex eigenvalues (in what follows, we
call the subspace $l=0$\ and the subspace $l\neq 0$\ the respective critical
and noncritical subspaces). This implies that there is the problem of
self-adjointness for the Dirac Hamiltonian with the magnetic-solenoid field,
at least in the critical subspace. We solve this problem in the subsequent
Sec.3.

\subsection{Solutions in 3+1 dimensions}

To take the symmetry of the problem under the $z$-translations into account,
we use the following representations for $\gamma $-matrices (see \cite{AMW89}%
), 
\begin{equation*}
\gamma ^{0}=\left( 
\begin{array}{cc}
\sigma ^{3} & 0 \\ 
0 & -\sigma ^{3}
\end{array}
\right) ,\;\gamma ^{1}=\left( 
\begin{array}{cc}
i\sigma ^{2} & 0 \\ 
0 & -i\sigma ^{2}
\end{array}
\right) ,\;\gamma ^{2}=\left( 
\begin{array}{cc}
-i\sigma ^{1} & 0 \\ 
0 & i\sigma ^{1}
\end{array}
\right) ,\;\gamma ^{3}=\left( 
\begin{array}{cc}
0 & I \\ 
-I & 0
\end{array}
\right) \,.
\end{equation*}
In $3+1$ dimensions, a complete set of commuting operators can be chosen as
follows ($\gamma ^{5}=-i\gamma ^{0}\gamma ^{1}\gamma ^{2}\gamma ^{3}$): 
\begin{equation}
H,\;P^{3}=-i\partial _{3},\;J^{3}=-i\partial _{\varphi }+\Sigma
^{3}/2,\;S^{3}=\gamma ^{5}\gamma ^{3}\left( M+\gamma ^{3}P^{3}\right) /M\,.
\label{abe3}
\end{equation}
We require that the wave function be an eigenvector for these operators, 
\begin{eqnarray}
H\Psi  &=&\varepsilon \Psi \,,  \label{abe4.1} \\
P^{3}\Psi  &=&p^{3}\Psi \,,  \label{abe4.2} \\
J^{3}\Psi  &=&j^{3}\Psi \,,  \label{abe4.3} \\
S^{3}\Psi  &=&s\widetilde{M}/M\Psi \,.  \label{abe4.4}
\end{eqnarray}
Here, $\widetilde{M}=\sqrt{M^{2}+(p_{3})^{2}}$, $p^{3}$ is the$\ z$%
-component of the momentum, and $j^{3}$ is the $z$-component of the total
angular momentum. We note that the energy eigenvalues can be positive, $%
\varepsilon =_{+}\varepsilon >0$ , or negative, $\varepsilon
={}_{-}\varepsilon <0$.$\;$The eigenvalues $j^{3}$ are half-integer, it is
convenient to use the representation $j^{3}=\left( l-l_{0}-\frac{1}{2}%
\right) $, where $l=0,\pm 1,\pm 2,...$. To specify the spin degree of
freedom, we choose the operator $S^{3}$ that is the $z$-component of the
polarization pseudovector \cite{ST68}, 
\begin{equation}
S^{0}=-\frac{1}{2M}\left( H\gamma ^{5}+\gamma ^{5}H\right) ,\;S^{i}=\frac{1}{%
2M}\left( H\Sigma ^{i}+\Sigma ^{i}H\right) \,,  \label{x.3}
\end{equation}
the eigenvalues of the\ corresponding spin projections are $s\widetilde{M}/M$%
, $s=\pm 1$.

Then, in $3+1$ dimensions, we can separate the spin and coordinate variables
and obtain the following representation for the spinors $\Psi $: 
\begin{eqnarray}
&&\Psi (x)=\exp \left\{ -i\varepsilon x^{0}+ip^{3}x^{3}\right\} \Psi
_{s}(x_{\perp })\,,  \notag \\
&&\Psi _{s}(x_{\perp })=N\left( 
\begin{array}{c}
\left[ 1+\left( p^{3}+s\widetilde{M}\right) /M\right] \psi _{\varepsilon
,s}(x_{\perp }) \\ 
\left[ -1+\left( p^{3}+s\widetilde{M}\right) /M\right] \psi _{\varepsilon
,s}(x_{\perp })
\end{array}
\right) \,.  \label{abe4}
\end{eqnarray}
Here, $\psi _{\varepsilon ,s}(x_{\perp })$ are two-component spinors, $%
x_{\perp }=\left( 0,x^{1},x^{2},0\right) $, and $N$ is a normalization
factor.

The equation (\ref{abe4.1}) is then reduced to the equation 
\begin{equation}
\left( \mathbf{\mbox{\boldmath$\sigma$\unboldmath}P}_{\perp }+s\widetilde{M}%
\sigma ^{3}\right) \psi _{\varepsilon ,s}(x_{\perp })=\varepsilon \psi
_{\varepsilon ,s}(x_{\perp }),\;P_{\perp }=\left( 0,P_{1},P_{2},0\right) \,.
\label{abe5}
\end{equation}
Representing $\psi _{\varepsilon ,s}\left( x_{\perp }\right) $ in the form 
\begin{equation}
\psi _{\varepsilon ,s}\left( x_{\perp }\right) =g_{l}(\varphi )\psi
_{l,s}\left( r\right) \,,  \label{abe5n0}
\end{equation}
where $g_{l}\left( \varphi \right) $\ is given by Eq. (\ref{abe11}), we
obtain the radial equation 
\begin{equation}
h_{s}\psi _{l,s}(r)=\varepsilon \psi _{l,s}(r),\;h_{s}=\Pi +s\widetilde{M}%
\sigma ^{3}\,,  \label{abe5n1}
\end{equation}
where $h_{s}$\ is the radial Hamiltonian acting on the subspace with the
spin quantum number $s$, $\Pi $\ is given by Eq. (\ref{abe13}). We note that 
\begin{equation}
\psi _{\varepsilon ,-1}(x_{\perp })=\sigma ^{3}\psi _{-\varepsilon
,1}(x_{\perp })\,.  \label{abe5b}
\end{equation}
We can see that for fixed $s$ and $p^{3}$, Eq. (\ref{abe5}) is similar to
Eq. (\ref{abe7a}) in $2+1$ dimensions. Therefore, after separating the
angular variable by (\ref{abe11}), the radial spinor $\psi _{l,+1}\left(
r\right) $ (\ref{abe5n0}) can be obtained from the radial spinor $\psi
_{l}\left( r\right) $ (\ref{abe11}) by substituting $M$ by $\widetilde{M}$.
The same is true for the particular case $l=0$. Here, the radial spinor\ $%
\psi _{\omega ,+1}\left( r\right) $ can be obtained from the radial spinor $%
\psi _{\omega }\left( r\right) $ (\ref{abe34}).

Using the results for the $\left( 2+1\right) $-dimensional case, we conclude
that in the critical subspace, the complex eigenvalues of Eq. (\ref{abe4.1})
do exist if $\mu \neq 0$. This means that the abovementioned problem of
self-adjointness for the Dirac Hamiltonian in $2+1$ dimensions is reproduced
in $3+1$ dimensions.

\section{Self-adjoint extensions}

We here solve this problem using von Neumann's theory of the  self-adjoint
extensions of symmetric operators \cite{RS72}. The canonical procedure
includes the following steps: initially defining the Dirac Hamiltonian $H$
as a symmetric operator, then evaluating its adjoint $H^{\dagger }$ and its
closure $\overline{H}$, then finding the deficiency subspaces $\mathcal{D}%
^{+}$ and $\mathcal{D}^{-}$\ and the deficiency indices, and, at last, in
the case of equal deficiency indices, describing the isometries from\ $%
\mathcal{D}^{+}$ to $\mathcal{D}^{-}$ that define the self-adjoint
extensions of $H$. This constitutes von Neumann's theory that is applicable
to the general case. In Sec. 5, we show that in our particular case, this
procedure can be significantly reduced: it is sufficient to evaluate $%
H^{\dagger }$ and then analyze its symmetry properties; of course, we thus
follow one of the main ideas of von Neumann.

\subsection{Extensions in 2+1 dimensions}

We first study the $\left( 2+1\right) $-dimensional case. This case was
partially considered in \cite{FP01}. The results of \cite{FP01} are
recovered by our consideration based on the detailed study of the Dirac
equation solutions. We also generalize these results for an arbitrary sign
of $B,$ which allows determining the nontrivial dependence of the spectrum
on the signs of $B$ and $\Phi $. The obtained results are then extended to
the $\left( 3+1\right) $-dimensional case. From the mathematical rigor
standpoint, our exposition is rather qualitative, some important technical
details can be found in Sec. 5.

The problem is to define the Hamiltonian (\ref{abe1}) as a \thinspace\
self-adjoint operator in the Hilbert space of square-integrable two-spinors $%
\psi \left( x_{\bot }\right) $. By separating variables, Eq. (\ref{abe11}),
the problem is reduced to the corresponding problem for the radial
Hamiltonian $h$ (\ref{abe12}), (\ref{abe13}) for every $l=0,\pm 1,...$ in
the Hilbert space of two-spinors $\psi \left( r\right) $ square integrable
on the half-line with the measure $rdr$. We start with the choice of the
initial\ domain $\mathcal{D}\left( h\right) $ for the operator $h$. Because
the source of the problem under consideration is the singularity of the AB
potential at the origin $r=0$, we first try to avoid the troubles associated
with this singularity. Therefore, let $\mathcal{D}\left( h\right) $ be the
(sub)space of absolutely continuous two spinors $\psi (r)$ vanishing
sufficiently fast as $r\rightarrow 0$; of course, $h\psi (r)$ must be square
integrable together with $\psi (r)$. It is easy to verify that the initial $h
$ is symmetric by integrating by parts. Its adjoint $h^{\dagger }$ is given
by the same expression (\ref{abe12}), (\ref{abe13}),\ but is defined on the
domain $\mathcal{D}(h^{\dagger })\supset \mathcal{D}(h)$ of absolutely
continuous two-spinors not necessarily vanishing as $r\rightarrow 0$ in the
critical subspace $l=0$ for $\mu \neq 0$. Its closure $\overline{h}$ is
defined on the domain $\mathcal{D}(\overline{h})$ of absolutely continuous
two-spinors vanishing as $r\rightarrow 0$.

Then we have to find the deficiency subspaces $\mathcal{D}^{+}$ and $%
\mathcal{D}^{-}$, $\mathcal{D}^{\pm }=$\textrm{Ker}$\left( h^{\dagger }\mp
iM\right) $ (here, $M$ is introduced by dimensional reasons), and the
deficiency indices $n_{\pm }\left( h\right) =\dim \left( \mathcal{D}^{\pm
}\right) $, i.e. to find the number of linearly independent
square-integrable solutions of the equations 
\begin{eqnarray}
&&h^{\dagger }\psi ^{\pm }(r)=\pm iM\psi ^{\pm }(r),\;\;h^{\dagger }=\Pi
^{\dagger }+\sigma ^{3}M\,,  \label{abe39} \\
&&\Pi ^{\dagger }=-i\left\{ \partial _{r}+\frac{\sigma ^{3}}{r}\left[ \mu +l-%
\frac{1}{2}\left( 1-\sigma ^{3}\right) +A\left( r\right) \right] \right\}
\sigma ^{1}\,.  \label{abe40}
\end{eqnarray}
In the noncritical subspaces $l\neq 0$, there are no such solutions. In the
critical subspace $l=0$, if $\mu \neq 0$, there is only one solution for $%
\psi ^{+}\left( r\right) $ and $\psi ^{-}\left( r\right) $, namely, 
\begin{eqnarray}
\psi ^{\pm }(r) &=&N\left( 
\begin{array}{l}
\phi _{1}(r) \\ 
\pm e^{\pm i\pi /4}\frac{\sqrt{\gamma }}{M}\phi _{-1}(r)
\end{array}
\right) ,\;B>0\,,  \label{abe41} \\
\psi ^{\pm }(r) &=&N\left( 
\begin{array}{l}
\phi _{1}(r) \\ 
\pm e^{\pm i\pi /4}\frac{M}{\sqrt{\gamma }}\phi _{-1}(r)
\end{array}
\right) ,\;B<0\,,  \label{abe42}
\end{eqnarray}
where 
\begin{equation*}
\phi _{\sigma }(r)=\psi _{\lambda ,\alpha }(\rho ),\;2\lambda
=-2M^{2}/\gamma -\xi \left( \mu -\left( 1-\sigma \right) /2\right) ,\;\sigma
=\pm 1\,,
\end{equation*}
see (\ref{abe25}), and $N$ is the normalization factor such that the
functions $\psi ^{\pm }(r)$ are normalized to unity. If $\mu =0,$ there are
no such solutions. This means that in the non-critical subspaces, the
deficiency indices are $(0,0)$, whereas in the critical subspace, these are $%
(1,1)$ unless $\mu =0;$ if $\mu =0,$ the deficiency indices are $(0,0)$.
Therefore, in the noncritical subspaces and, if $\mu =0,$ in the critical
subspace as well, the radial Hamiltonian $h$ is essentially self-adjoint,
i.e., its unique self-adjoint extension is its closure $\overline{h\text{ }}$%
\ $=h^{\dagger }$. This guaranties the completeness of the corresponding
solutions obtained in Sec. 2. In the critical subspace, the radial
Hamiltonian with $\mu \neq 0$ has a one-parameter family $\{h^{\Omega }\}$
of \thinspace\ self-adjoint extensions, which is homeomorphic to the group $%
U(1)$. Each member $h^{\Omega }$ of this family is defined by the isometry $%
\psi ^{+}(r)\rightarrow e^{i\Omega }\psi ^{-}(r)$ from $\mathcal{D}^{+}$
onto\ $\mathcal{D}^{-}$,$\;0\leq \Omega <2\pi $. The domain of $h^{\Omega }$
is 
\begin{equation}
\mathcal{D}\left( h^{\Omega }\right) =\left\{ \chi \left( r\right) =\psi
\left( r\right) +c\left[ \psi ^{+}\left( r\right) +e^{i\Omega }\psi
^{-}\left( r\right) \right] :\;\psi \left( r\right) \in \mathcal{D}(%
\overline{h})\right\} ,\;c\in \mathbb{C}\mathbf{\,},  \label{abe43n}
\end{equation}
We now note that $\mathcal{D}\left( h^{\Omega }\right) $ is completely
defined by the asymptotic boundary conditions on the two-spinors $\chi
\left( r\right) =(\chi ^{1}(r),\,\,\chi ^{2}(r))$ as $r\rightarrow 0$.
Namely, because $\psi \left( r\right) \rightarrow 0$ as $r\rightarrow $ $0$,
then, if $c\neq 0$, the behavior of $\chi \left( r\right) $ as $r\rightarrow
0$ is defined by the singular behavior of $\psi ^{+}(r)$ and $\psi ^{-}(r)$
near the origin. Using the behavior (\ref{4.147}) of the function $\psi
_{\lambda ,\alpha }(\rho )$ at small $\rho $, we find 
\begin{equation}
\lim_{r\rightarrow 0}\frac{\chi ^{1}\left( r\right) \left( Mr\right) ^{1-\mu
}}{\chi ^{2}\left( r\right) \left( Mr\right) ^{\mu }}=\left\{ 
\begin{array}{c}
\frac{i2^{1-\mu }\Gamma (1-\mu )\Gamma (\mu +M^{2}/\gamma )}{\left( \tan 
\frac{\Omega }{2}-1\right) \Gamma (\mu )\Gamma (1+M^{2}/\gamma )}\left( 
\frac{M^{2}}{\gamma }\right) ^{1-\mu },\;B>0 \\ 
\frac{i2^{1-\mu }\Gamma (1-\mu )\Gamma (1+M^{2}/\gamma )}{\left( \tan \frac{%
\Omega }{2}-1\right) \Gamma (\mu )\Gamma (1-\mu +M^{2}/\gamma )}\left( \frac{%
M^{2}}{\gamma }\right) ^{-\mu }\,,\;B<0
\end{array}
\right. \,  \label{abe43}
\end{equation}
if $c\neq 0.$ (Of course, $c\,\ $can be equal to zero, then $\chi
(r)\rightarrow 0$ as $r\rightarrow 0.$)$\,$ These alternative possibilities
are just the asymptotic boundary conditions that uniquely specify the domain 
$\mathcal{D}\left( h^{\Omega }\right) $ and thus the self-adjoint extension $%
h^{\Omega }$. The self-adjoint asymptotic boundary conditions can be
consolidated into a single formula 
\begin{equation}
\chi \left( r\right) =c\left( 
\begin{array}{c}
i\Lambda \left( Mr\right) ^{\mu -1} \\ 
\left( Mr\right) ^{-\mu }
\end{array}
\right) +\epsilon \left( r\right) \,,\;r\rightarrow 0\,,  \label{abe45}
\end{equation}
where $i\Lambda $ is given by the right hand side of (\ref{abe43}), $-\infty
\leq \Lambda \leq \infty $, $\epsilon \left( r\right) \rightarrow 0$ as $%
r\rightarrow 0,$ and $c$ is an arbitrary constant. We can verify that in the
limit $\gamma \rightarrow 0$, the right-hand sides of (\ref{abe43}) coincide
with the corresponding expressions obtained in \cite{G89} for the case of
pure AB field. In fact, the family of self-adjoint asymptotic boundary
conditions is the same as in pure AB-field case, and is independent from $B.$
These conditions are dictated by the singular behavior of the AB potential
at the origin.

For our purposes it is convenient to pass from the parametrization by $%
\Omega $ to the parametrization by the angle $\Theta $, $0\leq \Theta <2\pi $%
,\ such that 
\begin{equation}
\lim_{r\rightarrow 0}\frac{\chi ^{1}\left( r\right) \left( Mr\right) ^{1-\mu
}}{\chi ^{2}\left( r\right) \left( Mr\right) ^{\mu }}=i\tan \left( \frac{\pi 
}{4}+\frac{\Theta }{2}\right) \,  \label{abe44}
\end{equation}
if $c\neq 0$.

Therefore, the solutions (\ref{abe34}) obtained in Sec.2 must be subjected
to the asymptotic condition (\ref{abe44}) as $r\rightarrow 0,$ which
guaranties the orthogonality and completeness of the corresponding
solutions. Using (\ref{abe35}), (\ref{abe36}), and (\ref{4.147}), we find 
\begin{equation}
\tan \left( \frac{\pi }{4}+\frac{\Theta }{2}\right) =\left\{ 
\begin{array}{c}
-\frac{\left( \varepsilon +M\right) }{M}\,\frac{\Gamma (1-\mu )\Gamma (\mu
-\omega /2\gamma )}{2^{\mu }\Gamma (\mu )\Gamma (1-\omega /2\gamma )}\left( 
\frac{M^{2}}{\gamma }\right) ^{1-\mu },\;B>0\, \\ 
\frac{M}{\left( \varepsilon -M\right) }\,\frac{\Gamma (1-\mu )\Gamma
(1-\omega /2\gamma )}{2^{\mu -1}\Gamma (\mu )\Gamma (1-\mu -\omega /2\gamma )%
}\left( \frac{M^{2}}{\gamma }\right) ^{-\mu },\;B<0.
\end{array}
\right. \,.  \label{abe48}
\end{equation}

\subsection{Extensions in 3+1 dimensions}

We now pass to the $\left( 3+1\right) $-dimensional case. The helicity
operator $S_{h}=\mathbf{\Sigma P/|P}|$ is commonly used as the spin
operator. It is related to the zero-component of the polarization
pseudovector (\ref{x.3}) by $S_{h}=S^{0}M/\left| \mathbf{P}\right| $.
Extensions, in which the operator $S^{0}$ commutes with the Hamiltonian,
were constructed for the particular case $p^{3}=0$ \cite{CP94,ACP01}.
However, the set of such extensions does not exhaust all the possible
Hamiltonian extensions in $2+1$ dimensions. Our choice of the operator $%
S^{3} $ as the spin operator allows us to separate the spin variables from
the beginning and to remain with the extension problem for the radial
Hamiltonian only. Therefore, we can efficiently apply our experience in $2+1$
dimensions to the $\left( 3+1\right) $-dimensional case.

In the $\left( 3+1\right) $-dimensional case, after separating the $z$%
-variable, the Hamiltonian $H$ must be defined as the self-adjoint operator
in the Hilbert space\ $\mathcal{H}$ of four-spinors of the form (\ref{abe4}%
). This Hilbert space can be represented as the direct sum of two orthogonal
subspaces labelled by the spin quantum number $s$: $\mathcal{H}=\left\{ \Psi
_{+1}\right\} \oplus \left\{ \Psi _{-1}\right\} $. These subspaces are
invariant with respect to $H$, and consequently, the Hamiltonian (\ref{abe1}%
) can be independently considered in each of the subspaces. Using Eqs. (\ref
{abe5}) and (\ref{abe11}) allows reducing the problem to the radial
Hamiltonians $h_{s}$ (\ref{abe5n1}) acting on the subspaces of the
two-spinors with the given spin quantum number $s=\pm 1,...$ and orbital
quantum number $l=0,\pm 1,...$. The problem with the radial Hamiltonian $%
h_{s}\,$in the $(3+1)$-dimensional case is absolutely similar to the problem
with the radial Hamiltonian $h$ in the $(2+1)$-dimensional case and is
solved by the same von Neumann's method. Because the procedure for defining $%
h_{s}$ as the self-adjoint operator literally repeats the procedure for $h$
in the previous section, we only outline the main steps.

We first define the radial Hamiltonian $h_{s}$ as a symmetric operator
choosing for the initial domain $\mathcal{D}\left( h_{s}\right) $ the space
of absolutely continuous two-spinors vanishing at the origin. Its adjoint $%
h_{s}^{\dagger }$ and its closure $\overline{h_{s}}$ are described just in
the same terms as the respective $h^{\dagger \text{ }}$and $\overline{h\text{
}}$ from the previous subsection. We now apply von Neumann's theory to each
of the subspaces. To find the deficiency subspaces and deficiency indices of
the operators $h_{s}$, we have to solve the equations 
\begin{equation}
h_{s}^{\dagger }\psi _{s}^{\pm }\left( r\right) =\pm is\widetilde{M}\psi
_{s}^{\pm }\left( r\right) ,\;h_{s}^{\dagger }=\Pi ^{\dagger }+s\widetilde{M}%
\sigma ^{3},\;s=\pm 1\,,  \label{abe50x1}
\end{equation}
where $\Pi ^{\dagger }$ is given by (\ref{abe40}). These equations are the
copies of Eq. (\ref{abe39}). Using Eqs. (\ref{abe41}), (\ref{abe42}) and (%
\ref{abe5b}), we find that for $l=0$ and $\mu \neq 0,$ the solutions are 
\begin{eqnarray}
&&\psi _{s}^{\pm }(r)=N\left( 
\begin{array}{l}
\phi _{s,+1}(r) \\ 
\pm se^{\pm i\pi /4}\frac{\sqrt{\gamma }}{\widetilde{M}}\phi _{s,-1}(r)
\end{array}
\right) ,\;B>0\,,  \label{abe50x3} \\
&&\psi _{s}^{\pm }(r)=N\left( 
\begin{array}{l}
\phi _{s,+1}(r) \\ 
\pm se^{\pm i\pi /4}\frac{\widetilde{M}}{\sqrt{\gamma }}\phi _{s,-1}(r)
\end{array}
\right) ,\;B<0\,,  \label{abe50x4} \\
&&\phi _{s,\sigma }(r)=\psi _{\lambda ,\alpha }(\rho )\,,\;\alpha =\mu
-\left( 1+\sigma \right) /2\,,  \notag \\
&&2\lambda =-2\widetilde{M}^{2}/\gamma -\xi \left( \mu -\left( 1-\sigma
\right) /2\right) \,,\;\sigma =\pm 1,  \notag
\end{eqnarray}
whereas for $l\neq 0$ and for $l=0$ and $\mu =0$, there are no square
integrable solutions, which means that for each $s=\pm 1$, the deficiency
indices in the noncritical subspaces $l\neq 0$ are $(0,0)$, whereas in the
critical subspace $l=0$, the deficiency indices are $(1,1)$ unless $\mu =0$;
if $\mu =0$ these are $(0,0)$. This implies that in the noncritical
subspaces and, if $\mu =0$, in the critical subspace as well, the radial
Hamiltonian $h_{s}$ is essentially  self-adjoint, i.e., its unique 
self-adjoint extension is its closure $\overline{h\,}=h^{\dagger }$, whereas
in the critical subspace, if $\mu \neq 0,$ there exists a one-parameter
family $\{h_{s}^{\Omega _{s}}\}$ of the  self-adjoint extensions of the
radial Hamiltonian $h_{s}$, labelled by the parameter $\Omega _{s}\,,\;0\leq
\Omega _{s}<2\pi $. The domain of $h_{s}^{\Omega _{s}}$ is 
\begin{equation}
\mathcal{D}\left( h_{s}^{\Omega _{s}}\right) =\left\{ \chi _{s}\left(
r\right) =\psi _{s}\left( r\right) +c\left[ \psi _{s}^{+}\left( r\right)
+e^{i\Omega _{s}}\psi _{s}^{-}\left( r\right) \right] :\;\psi _{s}\left(
r\right) \in \mathcal{D}\left( \overline{h}_{s}\right) \right\} ,\;c\in 
\mathbb{C}\,.  \label{r12}
\end{equation}

Quite similarly to the $(2+1)$-dimensional case, $h_{s}^{\Omega _{s}}$ is
specified by the self-adjoint asymptotic boundary conditions on the
two-spinors $\chi (r)=(\chi ^{1}(r),\,\chi ^{2}(r))$ as $r\rightarrow 0$.
With the parametrization by the angle $\Theta _{s}$ similar to (\ref{abe44}%
), these boundary conditions are 
\begin{equation}
\lim_{r\rightarrow 0}\frac{\chi _{s}^{1}\left( r\right) \left( \widetilde{M}%
r\right) ^{1-\mu }}{\chi _{s}^{2}\left( r\right) \left( \widetilde{M}%
r\right) ^{\mu }}=si\tan \left( \frac{\pi }{4}+\frac{\Theta _{s}}{2}\right)
,\;s=\pm 1  \label{r14}
\end{equation}
if $c\neq 0$ and $\chi _{s}(r)\rightarrow 0$ as $r\rightarrow 0$ if $c=0,$
which can be consolidated into a single formula 
\begin{equation*}
\chi _{s}\left( r\right) =c_{s}\left( 
\begin{array}{c}
i\Lambda _{s}\left( \tilde{M}r\right) ^{\mu -1} \\ 
\left( \tilde{M}r\right) ^{-\mu }
\end{array}
\right) +\epsilon _{s}\left( r\right) \,,\;r\rightarrow 0\,,
\end{equation*}
where $i\Lambda _{s}$ is given by the right hand side of (\ref{r14}), $%
-\infty \leq \Lambda _{s}\leq \infty $, $\epsilon \left( r\right)
\rightarrow 0$ as $r\rightarrow 0,$ and $c_{s}$ are arbitrary constants.

Therefore, in each subspace$\;s=\pm 1$ the above solutions $\psi _{\omega
,s}\left( r\right) $ in the critical subspace $l=0$ must be subjected to the
condition (\ref{r14}).

The remarks on the orthogonality and completeness of the corresponding
solutions are similar to those in previous Sec. 3.1.

The final result is that in 3 + 1 dimensions, there is the two-parameter
family of the self-adjoint Dirac Hamiltonians. This family is the manifold $%
U(1)\times U(1)$. We emphasize that this is true only if we require that $%
S^{3}$ be conserved. If we don't require the conservation of $S^{3}$, then
it follows from the above consideration (for example, any vector satisfying
Eqs. (\ref{abe4.1})-(\ref{abe4.3}) can be constructed using two orthogonal
vectors determined as a superposition of $\Psi _{+1}$ and $\Psi _{-1}$) that
the deficiency indices in the critical subspace are $\left( 2,2\right) $
unless $\mu =0$. Consequently, the manifold of self-adjoint Dirac
Hamiltonians is the group $U(2)$ and is four-parametric. Only for the
submanifold $U(1)\times U(1)\subset U(2)$, the Hamiltonian $H$ and the
operator $S^{3}$ have a common set of eigenfunctions.

\subsection{Spectra of  self-adjoint extensions}

We now study the spectra of the  self-adjoint extensions $h^{\Omega }$. To
find these, we have to solve the transcendental equations (\ref{abe48}) for $%
\omega $ considering two branches of $\varepsilon $, one for particles and
another one for antiparticles, $_{\pm }\varepsilon =\pm \sqrt{M^{2}+\omega }$%
. Introducing the notation 
\begin{eqnarray}
\,\omega  &=&2\gamma x,\;\;x=\,_{\varsigma }x=\left( \,_{\varsigma
}\varepsilon ^{2}-M^{2}\right) /2\gamma \,,\;Q\left( x\right) =\frac{%
\varepsilon }{M}+1\,,\;\varsigma =\pm \,,  \notag \\
\,\,\eta  &=&\frac{2^{\mu }\Gamma (\mu )}{\Gamma (1-\mu )}\tilde{\eta}\left(
\mu \right) ,\;\;\tilde{\eta}\left( \mu \right) =-\tan \left( \frac{\pi }{4}+%
\frac{\Theta }{2}\right) \left( \frac{\gamma }{M^{2}}\right) ^{1-\mu }\,,
\label{abe50n}
\end{eqnarray}
for $B>0$, we can rewrite Eq. (\ref{abe48}) as 
\begin{equation}
Q\left( _{\varsigma }x\right) \frac{\Gamma (\mu -\,_{\varsigma }x)}{\Gamma
(1-\,_{\varsigma }x)}=\eta \,.  \label{abe501}
\end{equation}
Given $\omega $ for $B>0$, we can obtain $\omega $ for $B<0$ with the
substitutions 
\begin{equation*}
\varsigma \rightarrow -\varsigma ,{\LARGE \ \;}\tilde{\eta}\left( \mu
\right) \rightarrow 1/\tilde{\eta}\left( \mu \right) ,\;\;\mu \rightarrow
1-\mu \,.
\end{equation*}
In what follows, we therefore consider the case $B>0$ only.

The possible solutions $x=x\left( \eta \right) $ of the equation (\ref
{abe501}) are the functions of the parameter $\eta $ (i.e., of $\mu ,$ $%
\gamma /M^{2}$, and $\Theta $) and are labelled by $m=0,1,...$ . We find the
following asymptotic representations for these solutions as $\left| \eta
\right| \rightarrow 0$\thinspace : 
\begin{eqnarray}
&&x_{m}\left( \eta \right) =m+\Delta x_{m},\;\;\Delta x_{m}=\frac{\sin
\left( \pi \mu \right) \Gamma (m+1-\mu )}{\pi \Gamma (m)Q\left( m\right) }%
\eta \,,\;\;m=1,2,3,\ldots ,  \notag \\
&&_{-}x_{0}\left( \eta \right) =-\frac{\eta M^{2}}{\gamma \Gamma \left( \mu
\right) }\;.  \label{abe502}
\end{eqnarray}
All $x_{m}\left( 0\right) ,\;m=1,2,...$ are positive and integer. The
asymptotic representation of $_{+}x_{0}\left( \eta \right) $ as $\left| \eta
\right| \rightarrow 0$ is discussed below. The function $_{+}x_{0}\left(
\eta \right) $ vanishes at the point $\eta =2\Gamma (\mu )$ , and in the
neighborhood of this point, it has the form 
\begin{equation}
\,_{+}x_{0}\left( \eta \right) =\frac{\Gamma (\mu )-\eta /2}{\Gamma (\mu
)\left( \psi (\mu )-\psi (1)\right) }\,.  \label{abe504}
\end{equation}
Here, $\psi (x)$ is the logarithmic derivative of the gamma function $\Gamma
(x)$, and $-\psi (1)\simeq 0.577\,$ is the Euler-Mascheroni constant \cite
{HTF1}. As $\left| \eta \right| \rightarrow \infty $,we find the following
asymptotic representations: 
\begin{eqnarray}
&&_{\varsigma }x_{m}\left( \varsigma \eta \right) =m+\mu +\Delta
x_{m},\;\;m=0,1,2,\ldots \;,{}\;\eta \rightarrow \infty \,,  \notag \\
&&_{\varsigma }x_{m}\left( \varsigma \eta \right) =m-1+\mu +\Delta
x_{m},\;\;m=1,2,3,\ldots \;,{}\;\eta \rightarrow -\infty \,,  \notag \\
&&\Delta x_{m}=-\frac{\sin \left( \pi \mu \right) \Gamma (m+\mu )Q\left(
m+\mu \right) }{\pi \Gamma (m+1)\eta }\;.  \label{abe505}
\end{eqnarray}
These approximations hold only for $\left| \Delta x_{m}\right| $ $\ll \mu \,$%
and $\left| x_{0}\left( \eta \right) \right| \ll \mu $.

According to\ \cite{W80} (see Corollary 1 of Theorem 8.19 therein), if $T_{1}
$ and $T_{2}$ are two  self-adjoint extensions of the same symmetric
operator with equal finite deficiency indices $(d,d),$ then any interval $%
\left( a,b\right) \subset \mathcal{R}$ not intersecting the spectrum of $%
T_{1}$ contains only isolated eigenvalues of the operator $T_{2}$ with total
multiplicity at most $d$. We take the extension $h^{\Omega }$ with $\Theta
=\pi /2$ whose eigenvalues are $\,_{+}\varepsilon =M\sqrt{1+2\gamma
\,_{+}x_{0}\left( \infty \right) /M^{2}}$\ and $\,_{\pm }\varepsilon =\pm M%
\sqrt{1+2\gamma \,_{\pm }x_{m}\left( \pm \infty \right) /M^{2}}$, $m\geq 1$.
The above theorem implies that if $\left( a,b\right) $ is an open interval,
where$\ a\,$\ and $b$ are two subsequent eigenvalues of $h^{\Omega }$ with $%
\Theta =\pi /2$, or $_{\pm }\varepsilon =0$, then any  self-adjoint
extension $h^{\Omega }$ at $\Theta \neq \pi /2$\ has at most one eigenvalue
in $\left( a,b\right) $. According to\ \cite{AG81} (see Theorem 3 from Sec.
105 of Chapter VIII therein ), for any $\varepsilon \in \left( a,b\right) $,
there exists a  self-adjoint extension $h^{\Omega }$ with the eigenvalue $%
\varepsilon $. As it follows from (\ref{abe501}) and (\ref{abe505}), in the
ranges $(m-1+\mu \leq \,_{\pm }x_{m}\left( \eta \right) \leq m+\mu $, $m\geq
1)$ and $(-M^{2}/2\gamma \leq _{+}x_{0}\left( \eta \right) \leq \mu )$, the
functions $_{\pm }x\left( \eta \right) =\left( \,_{\pm }\varepsilon
^{2}-M^{2}\right) /2\gamma $ are one-valued and continuous. This observation
is in complete agreement with the above general Theorems. The functions $%
\,_{\pm }x_{m}\left( \eta \right) $ were found numerically in the weak
field, $\gamma /M^{2}\ll 1$, for some first $m$'s. The plots of these
functions (for $\mu =0.8$) are presented in Figs. 1 and 2.

\begin{figure}[th]
\centering
\includegraphics[width=3.5 in]{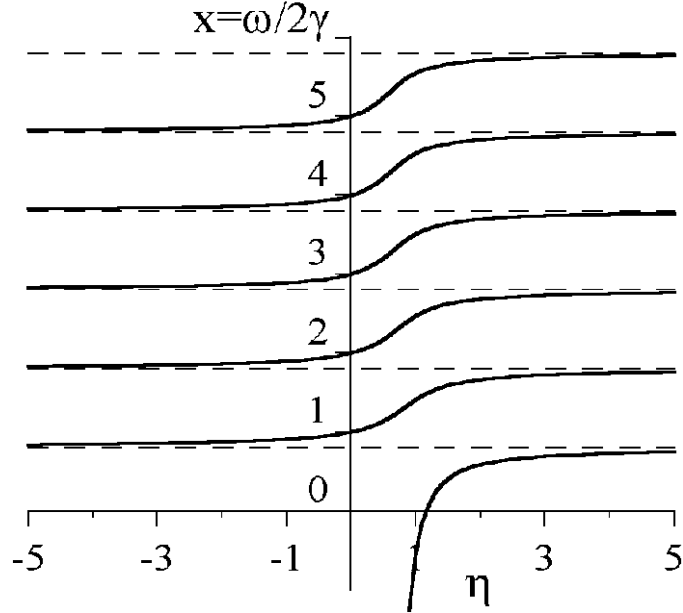}
\caption{Particle lower energy levels in dependence on the parameter $%
\protect\eta _{+}=\frac{\Gamma \left( \protect\mu \right) }{\Gamma \left( 1-%
\protect\mu \right) }\left( \frac{\protect\gamma }{2M^{2}}\right) ^{1-%
\protect\mu }\tan \left( \frac{\protect\pi }{4}+\frac{\Theta }{2}\right) $ }
\end{figure}

\begin{figure}[th]
\centering 
\includegraphics[width=3.5 in]{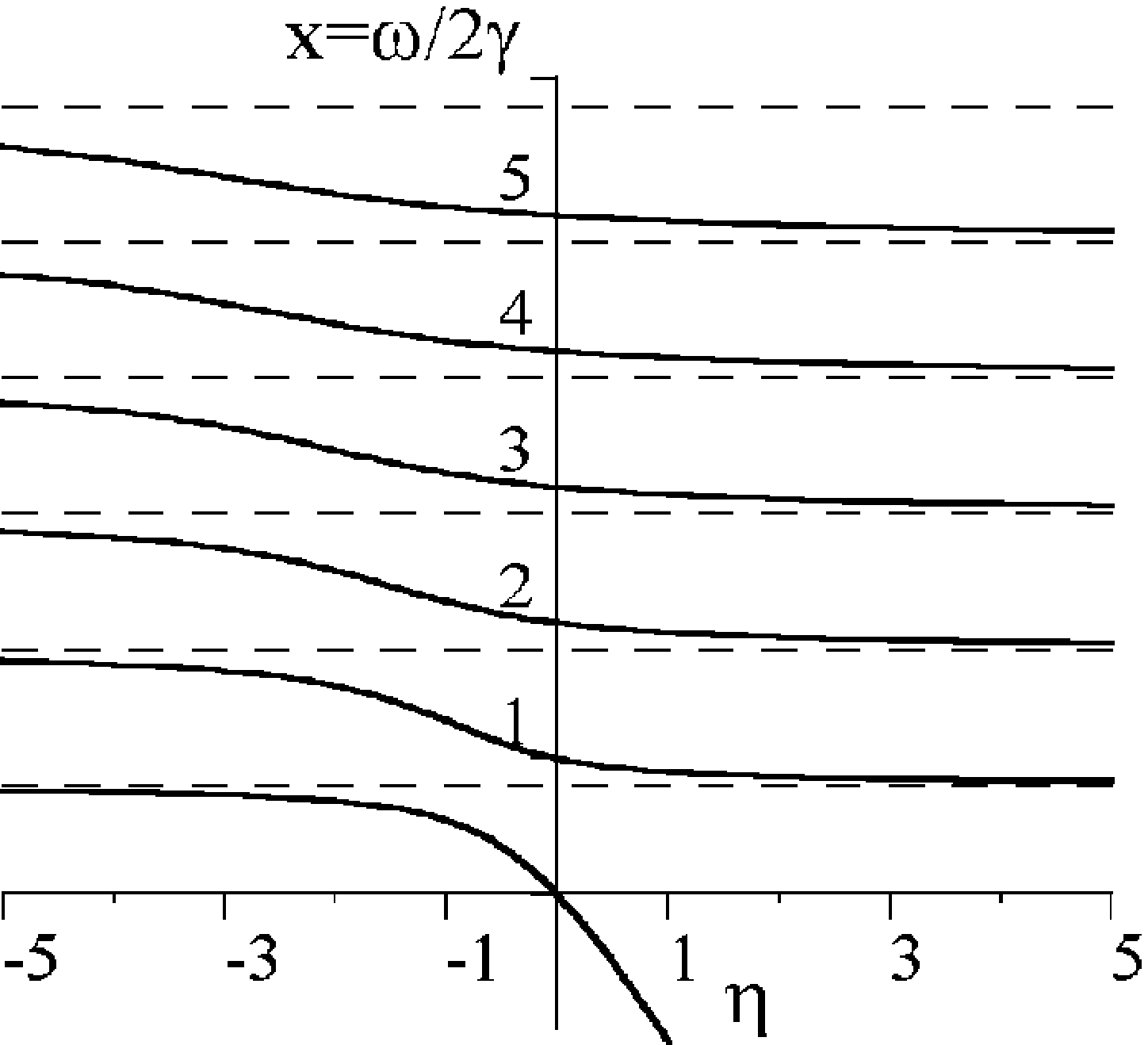}
\caption{Antiparticle lower energy levels in dependence on the parameter $%
\protect\eta _{-}=\frac{\Gamma \left( \protect\mu \right) }{\Gamma \left( 1-%
\protect\mu \right) }\left( \frac{\protect\gamma }{2M^{2}}\right) ^{-\protect%
\mu }\tan \left( \frac{\protect\pi }{4}+\frac{\Theta }{2}\right) $}
\end{figure}

We can see that $\delta x_{m}=x_{m+1}\left( \eta \right) -x_{m}\left( \eta
\right) \rightarrow 1$ with increasing $m$. It follows from the equation (%
\ref{abe501}) that 
\begin{equation}
\delta x_{m}-1=\pi ^{-1}\left\{ \cot \left( \pi x_{m}\right) -\cot \left[
\pi \left( x_{m}-\mu \right) \right] \right\} ^{-1}\left( \frac{1-\mu }{x_{m}%
}-\delta Q\right) ,\;m\gg 1\,,  \label{abe506}
\end{equation}
where $\delta Q=\left. \frac{d}{dx}\ln Q(x)\right| _{x=x_{m}}\leq 1/x_{m}\,$%
. The curve $x_{5}\left( \eta \right) $ can give an idea about the behavior
of the functions $x_{m}\left( \eta \right) $ for large $m\,$.

In the sequel, we discuss some limiting cases.

We first consider the weak fields $B$ for which $\gamma /M^{2}\ll 1$ and the
nonrelativistic electron energies, $x_{m}\left( \eta \right) \gamma
/M^{2}\ll 1$. Here, the functions $\,_{\pm }x\left( \eta \right) $ change
significantly in the neighborhood of $\eta =0$ only. Beyond the neighborhood
of $\eta =0,$ the functions $\,_{\pm }x\left( \eta \right) $ take the values
close to the corresponding asymptotic values given by (\ref{abe505}).

In the ultrarelativistic case, $x_{m}\left( \eta \right) \gamma /M^{2}\gg 1$%
, the behavior of $x_{m}\left( \eta \right) $ qualitatively depends on $\mu
\,$. One can distinguish the three cases: $0<\mu <1/2$, $\mu >1/2$, and $\mu
=1/2$. If $0<\mu <1/2,$ then the interval near $\eta =0$ on which the
functions change significantly narrows down with increasing $m$. If $\mu
>1/2,$ then this interval expands with increasing $m$ . For $\mu =1/2$ and $-%
\frac{1}{2}<\left( \frac{1}{4}+\frac{\Theta }{2\pi }\right) <\frac{1}{2}$,
we find the asymptotic representation 
\begin{equation}
\,_{\varsigma }x_{m}\left( \eta \right) =m+\varsigma \left( \frac{1}{4}+%
\frac{\Theta }{2\pi }\right) ,{}\;m\gg 1\;.  \label{abe506a}
\end{equation}

We see that negative $_{\pm }x_{0}\left( \eta \right) $ exist only for $\eta
>0$. This means that in the problem under consideration, there exist only
one particle state and only one antiparticle state with energies $\left|
\varepsilon \right| <M$ for $\pi /2<\Theta <3\pi /2$. The same situation was
observed in the pure AB field case \cite{G89}. The minimal admissible
negative $x_{0}\left( \eta \right) $ is defined by the condition $%
\varepsilon =0$. For the strong fields $B$ for which $\gamma /M^{2}\sim 1$,
the quantity $x_{0}\left( \eta \right) $\ is close to zero. Let $\Theta _{0}$
correspond to such an extension that admits $\varepsilon =0$. The value of $%
\Theta _{0}$ is defined by 
\begin{equation}
\tan \left( \frac{\pi }{4}+\frac{\Theta _{0}}{2}\right) =-\frac{\Gamma
(1-\mu )\Gamma (\mu +M^{2}/2\gamma )}{2^{\mu }\Gamma (\mu )\Gamma
(1+M^{2}/2\gamma )}\left( \frac{M^{2}}{\gamma }\right) ^{1-\mu }.
\label{abe507}
\end{equation}
In the weak fields, $\gamma /M^{2}\ll 1$, $x_{0}\left( \eta \right) $ takes
large absolute values, and the angle $\Theta _{0}$ is defined by 
\begin{equation}
\tan \left( \frac{\pi }{4}+\frac{\Theta _{0}}{2}\right) =-\frac{\Gamma
(1-\mu )}{2^{2\mu -1}\Gamma (\mu )}\,  \label{abe508}
\end{equation}
and independent from the magnetic field. It follows from (\ref{abe507}) that
in the superstrong fields $B$ for which $\gamma /M^{2}\gg 1$, the angle $%
\Theta _{0}$ is also independent from the magnetic field.

In the weak magnetic fields, $\gamma /M^{2}\ll 1$, and for the
nonrelativistic energies, $x_{0}\gamma /M^{2}\ll 1$, we obtain the relations 
\begin{eqnarray}
&&\,_{+}x_{0}\left( \eta \right) =-\left( 2/\eta \right) ^{1/\left( 1-\mu
\right) }\,,  \label{abe509} \\
&&\,_{-}x_{0}\left( \eta \right) =-\left( \eta M^{2}/\gamma \right) ^{1/\mu
}\,  \label{abe5010}
\end{eqnarray}
that are valid if $\eta $\ in (\ref{abe509}) is small and $\eta M^{2}/\gamma
\gg 1$ in (\ref{abe5010}).

We now consider the particular case $\Theta =-\pi /2.$ It follows from (\ref
{abe48}) that for$\ B>0$, there exists $_{-}\varepsilon =-M$. The energies $%
\left| \varepsilon \right| >M$ are defined by the poles of $\Gamma (1-x)$ or 
$\Gamma (1-\mu -x)$ for $B>0$ or $B<0$, respectively. The spectrum $%
\varepsilon $ coincides with the one defined by Eqs. (\ref{abe30}) and (\ref
{abe31}) for $\psi ^{I}$. Moreover, using the relation (\ref{4.140}), we can
see that the spinors $\psi _{\omega }(r)$ coincide with $\psi ^{I}$ up to a
normalization constant, 
\begin{equation}
\psi _{\omega }\left( r\right) \propto \psi ^{I}\left( r\right) \;\mathrm{%
for\;}\Theta =-\pi /2\,.  \label{abe50a}
\end{equation}

In the case $\Theta =\pi /2$, we have the following picture: It follows from
(\ref{abe48}) that for $B<0$, there exists $_{+}\varepsilon =M$. The
energies $\left| \varepsilon \right| >M$ are defined by the poles of $\Gamma
(\mu -x)$ or $\Gamma (1-x)$ for $B>0$ or $B<0$, respectively. The spectrum $%
\varepsilon $ coincides with the one given by Eqs. (\ref{abe30}) and (\ref
{abe31})\ for $\psi ^{II}$. It follows from (\ref{4.140}) that the spinor $%
\psi _{\omega }(r)$ coincides with $\psi ^{II}$ up to a normalization
constant, 
\begin{equation}
\psi _{\omega }(r)\propto \psi ^{II}\left( r\right) \;\mathrm{for\;}\Theta
=\pi /2\,.  \label{abe50b}
\end{equation}

Using the results for $B<0$, which are presented in Appendix B, we can
conclude that the spectrum asymmetry holds for the spinning particles in the
magnetic-solenoid field. There is a relation between the three-dimensional
chiral anomaly and the fermion zero modes in a uniform magnetic field\ \cite
{NS83} (for review, see \cite{J82,NS86}). We see that the effect also holds
in the presence of the AB potential.

The spectrum asymmetry is known in 2+1 QED with the uniform magnetic field.
In the uniform magnetic field, the states with $\omega =0$ for $l\neq 0$ are
observed if \textrm{sgn}$l=-\mathrm{sgn}B$ (for the antiparticle if $B>0$
and for the particle if $B<0$). The spectrum changes mirror-like with the
change of the sign of the magnetic field. We see that for $l\neq 0$, the
spectrum properties in the magnetic-solenoid field is similar to the
spectrum properties in the uniform magnetic field. The presence of the AB
potential is especially essential for the states with $l=0$ where the
particle penetrates the solenoid.

The spectra in $3+1$ dimensions can be obtained from the results in $2+1$
dimensions. Namely, we use the fact that the solutions $\psi _{\varepsilon
,1}(x_{\perp })$ in $3+1$ dimensions are obtained from the solutions $\psi
_{\varepsilon }^{(1)}(x_{\perp })$ in $2+1$ dimensions. Therefore, the
spectra in $3+1$ dimensions are obtained from the results in $2+1$
dimensions by substituting $M$ by $\widetilde{M}$ and using the relation (%
\ref{abe5b}). As a consequence, we obtain an additional interpretation of
Figs. 1 and 2. In particular, Fig.1 shows the lower energy levels for
particles with spin $s=1$, and Fig. 2 shows the lower energy levels for
particles with spin $s=-1$.

\section{Solenoid regularization}

\subsection{Spinning particle case}

We can introduce the AB field as a limiting case of a finite-radius solenoid
field (the regularized AB field), which allows fixing the extension
parameters. This approach to the pure AB field was first proposed by Hagen 
\cite{H90}. In what follows, we consider the problem in the presence of a
uniform magnetic field. For this, we have to study the solutions of the
Dirac equation (\ref{abe1}) with the combination of a finite-radius solenoid
field and a collinear uniform magnetic field.

Let the solenoid have a radius $R$. We assume that inside the solenoid,
there is an axially symmetric magnetic field $B^{in}(r)$ that creates the
flux $\Phi =\left( l_{0}+\mu \right) \Phi _{0},\;\Phi _{0}=2\pi /e$, such
that $e\int_{0}^{R}B^{in}\left( r\right) rdr=l_{0}+\mu $, and outside the
solenoid ($r>R$), the field $B^{in}(r)$ vanishes.. The function $B^{in}(r)$
is arbitrary but such that the integrals in the functions $\vartheta \left(
x\right) \;$and $b\left( x\right) $ in (\ref{sol-inside}) are convergent. We
take the potentials of the field $B^{in}(r)$ in the form 
\begin{equation}
eA_{1}^{in}=\vartheta \left( x\right) \frac{\sin \varphi }{Rx}%
,\;eA_{2}=-\vartheta \left( x\right) \frac{\cos \varphi }{Rx}\,,
\label{abe52}
\end{equation}
where 
\begin{equation*}
\;\vartheta \left( x\right) =\int_{0}^{x}f\left( x^{\prime }\right)
x^{\prime }dx^{\prime },\;f(x)=R^{2}eB^{in}(xR),\;x=r/R\;.
\end{equation*}
The potentials of the uniform magnetic field are 
\begin{equation}
A_{0}=0,\;A_{1}=A\left( r\right) \frac{\sin \varphi }{r},\;A_{2}=-A\left(
r\right) \frac{\cos \varphi }{r}\;,\;\;A\left( r\right) =Br^{2}/2\;.
\label{x.1}
\end{equation}
Outside the solenoid, the potentials have form (\ref{abe2}).

We analyze the solutions of the Dirac equation in the field defined above .
For this, we have to solve the equation inside and outside the solenoid and
continuously sew the corresponding solutions. We call the corresponding
Dirac spinors the respective inside and outside solutions.

We first study the problem in $2+1$ dimensions. We require that the
solutions be square integrable and regular as $r\rightarrow 0$. In the same
way as in Sec. 2, we obtain that the inside radial spinors $\psi _{\omega
,l}^{in}(r)$\ ($r\leq R$) satisfy the equation 
\begin{equation*}
h^{in}\psi _{\omega ,l}^{in}\left( r\right) =\varepsilon \psi _{\omega
,l}^{in}\left( r\right) ,\;\;h^{in}=\Pi ^{in}+\sigma ^{3}M\;,
\end{equation*}
where 
\begin{equation}
\Pi ^{in}=-\frac{i}{R}\left\{ \partial _{x}+\frac{\sigma ^{3}}{x}\left[
l-l_{0}-\frac{1}{2}\left( 1-\sigma ^{3}\right) +\vartheta \left( x\right)
+\xi \rho _{R}x^{2}\right] \right\} \sigma ^{1},\;\rho _{R}=\gamma R^{2}/2\;.
\label{abe55}
\end{equation}
We require that the functions $\psi _{\omega ,l}^{in}\left( r\right) $ be
square integrable on the interval $\left( 0,R\right) $. For $\omega =0$ ($%
\left| \varepsilon \right| =M$), the solutions are 
\begin{eqnarray}
_{+}\psi _{0,l}^{in}(r) &=&\phi _{0,l,1}^{in}(x)\upsilon
_{1}\;,{}\;l-l_{0}\geq 1\;,  \notag \\
_{-}\psi _{0,l}^{in}(r) &=&\phi _{0,l,-1}^{in}(x)\upsilon
_{-1}\;,{}\;l-l_{0}\leq 0\;,  \notag \\
\phi _{0,l,\sigma }^{in}(x) &=&cx^{|\eta |}\exp \left\{ \sigma \int_{0}^{x}%
\tilde{x}^{-1}\left( \vartheta (\tilde{x})+\xi \rho _{R}\tilde{x}^{2}\right)
d\tilde{x}\right\} ,\;\eta =l-l_{0}-\left( 1+\sigma \right) /2\;,
\label{abe56}
\end{eqnarray}
where $c$ is an arbitrary constant. For $\omega \neq 0$, we represent the
spinors in the form 
\begin{equation*}
\psi _{\omega ,l}^{in}\left( r\right) =\left( 
\begin{array}{c}
\psi _{1}^{in}(r) \\ 
\psi _{2}^{in}(r)
\end{array}
\right) =\left[ \sigma ^{3}\left( \varepsilon -\Pi ^{in}\right) +M\right] 
\left[ c_{1}\phi _{l,1}^{in}(x)\upsilon _{1}+ic_{-1}\phi
_{l,-1}^{in}(x)\upsilon _{-1}\right] \,,
\end{equation*}
where $c_{\sigma }$ are arbitrary constants. The functions $\phi _{l,\sigma
}^{in}\left( x\right) $ satisfy the equation 
\begin{equation}
\left[ \frac{1}{x}\frac{\partial }{\partial x}x\frac{\partial }{\partial x}-%
\frac{1}{x^{2}}\left( \eta +\vartheta \left( x\right) +\xi \rho
_{R}x^{2}\right) ^{2}+\omega R^{2}-\sigma \left( f\left( x\right) +2\xi \rho
_{R}\right) \right] \phi _{l,\sigma }^{in}\left( x\right) =0  \label{abe58}
\end{equation}
and must be regular at $r=0$ in order to satisfy the square integrability
condition for $\psi _{\omega ,l}^{in}\left( r\right) $. Our prime interest
is in the limiting case $R\rightarrow 0$. For our purposes, it is enough to
use the approximation $\rho _{R}\ll 1$, $\omega R^{2}\ll 1$. Rejecting the
terms proportional to $R^{2}$ in (\ref{abe55}) and (\ref{abe58}), we find
that solutions of Eq. (\ref{abe58}) are 
\begin{eqnarray}
&&\phi _{l,\sigma }^{in}\left( x\right) =\left\{ 
\begin{array}{c}
cx^{|\eta |}e^{\sigma b(x)},\;\sigma \eta \geq 0\,, \\ 
cx^{-|\eta |}e^{\sigma b(x)}\int_{0}^{x}d\tilde{x}\tilde{x}^{2|\eta
|-1}e^{-2\sigma b(\tilde{x})},\;\sigma \eta <0\,,
\end{array}
\right.  \notag \\
&&b(x)=\int_{0}^{x}d\tilde{x}\tilde{x}^{-1}\vartheta (\tilde{x})\,.
\label{sol-inside}
\end{eqnarray}

The outside solutions ($r\geq R$) satisfy the equation 
\begin{equation}
h\psi _{\omega ,l}^{out}\left( r\right) =\varepsilon \psi _{\omega
,l}^{out}\left( r\right)  \label{abe57}
\end{equation}
and must be square integrable on the interval $\left( R,\infty \right) $.
Here, $h$ is defined by Eqs. (\ref{abe12}) and (\ref{abe13}). The general
form of the outside solutions is 
\begin{eqnarray}
&&\psi _{\omega ,l}^{out}\left( r\right) =\left[ \sigma ^{3}\left(
\varepsilon -\Pi \right) +M\right] \left( c_{1}\phi _{l,1}^{out}(r)\upsilon
_{1}+ic_{-1}\phi _{l,-1}^{out}(r)\upsilon _{-1}\right) \,,  \notag \\
&&\phi _{l,\sigma }^{out}\left( r\right) =\psi _{\lambda ,\alpha }\left(
\rho \right) \;,\alpha =l+\mu -\left( 1+\sigma \right) /2,\;2\lambda =\omega
/\gamma -\xi \left( l+\mu -\left( 1-\sigma \right) /2\right) \,.
\label{s-out}
\end{eqnarray}
The solutions $\psi _{\omega ,l}^{out}\left( r\right) $ and $\psi _{\omega
,l}^{in}\left( r\right) $ must be sewed continuously at $r=R$, 
\begin{equation}
\psi ^{out}\left( R\right) =\psi ^{in}\left( R\right) ,  \label{bc-out-in}
\end{equation}
and must satisfy the normalization condition 
\begin{eqnarray}
&&N_{\omega ,l}^{in}+N_{\omega ,l}^{out}=1\,,  \notag \\
N_{\omega ,l}^{in} &=&\int_{0}^{R}\left( \psi _{\omega ,l}^{in}(r)\right)
^{\dagger }\psi _{\omega ,l}^{in}(r)rdr\;,\;\;N_{\omega
,l}^{out}=\int_{R}^{\infty }\left( \psi _{\omega ,l}^{out}(r)\right)
^{\dagger }\psi _{\omega ,l}^{out}(r)rdr\,.  \label{abe611}
\end{eqnarray}
We can treat the AB field as a limiting case of the finite-radius solenoid
field if 
\begin{equation}
\lim_{\rho _{R}\rightarrow 0}N_{\omega ,l}^{in}=0\,.  \label{x.2}
\end{equation}
We can realize the sewing condition (\ref{bc-out-in}) imposing the following
conditions on the functions $\phi _{l,\sigma }^{in}\left( r\right) $\ and $%
\phi _{l,\sigma }^{out}\left( r\right) $ at $r=R$: 
\begin{equation}
\phi \left( R-\epsilon \right) =\phi \left( R+\epsilon \right) ,\;\frac{d}{dr%
}\phi \left( R-\epsilon \right) =\frac{d}{dr}\phi \left( R+\epsilon \right)
\,.  \label{cond-cont}
\end{equation}

It is convenient to use the representation (\ref{4.140}) for $\psi _{\lambda
,\alpha }\left( \rho \right) $ in (\ref{s-out}).The functions $\phi
_{l,\sigma }^{out}\left( r\right) $ then are 
\begin{eqnarray}
&&\phi _{l,\sigma }^{out}\left( r\right) =a_{\sigma }I_{n_{\sigma
},m_{\sigma }}\left( \rho \right) +b_{\sigma }I_{m_{\sigma },n_{\sigma
}}\left( \rho \right) ,\;n_{\sigma }=\lambda -\frac{1-\alpha }{2}%
,\;m_{\sigma }=\lambda -\frac{1+\alpha }{2}\,,  \notag \\
&&a_{\sigma }=K\sin n_{\sigma }\pi ,\;b_{\sigma }=-K\sin m_{\sigma }\pi ,\;K=%
\frac{\sqrt{\Gamma \left( 1+n_{\sigma }\right) \Gamma \left( 1+m_{\sigma
}\right) }}{\sin \left( n_{\sigma }-m_{\sigma }\right) \pi }\,,
\label{abe59}
\end{eqnarray}
where $n_{\sigma }$, $m_{\sigma }$ are real numbers.

Using (\ref{cond-cont}), we can find the coefficients $a_{\sigma
},\;b_{\sigma }$: for the case $l-l_{0}\leq 0$, 
\begin{eqnarray}
&&a_{1}=\rho _{R}^{-\left( l+\mu -1\right) /2}c\tilde{a}_{1},{}\;b_{1}=\rho
_{R}^{\left( l+\mu -1\right) /2}c\tilde{b}_{1}\,,  \label{abe60a} \\
&&a_{-1}=\rho _{R}^{-\left( l+\mu \right) /2+1}c\tilde{a}_{-1},{}\;b_{-1}=%
\rho _{R}^{\left( l+\mu \right) /2}c\tilde{b}_{-1}\,,  \label{abe60b}
\end{eqnarray}
whereas for the case $l-l_{0}>0$, 
\begin{eqnarray}
&&a_{1}=\rho _{R}^{-\left( l+\mu -1\right) /2}c^{\prime }\tilde{a}%
_{1}^{\prime },{}\;b_{1}=\rho _{R}^{\left( l+\mu -1\right) /2+1}c^{\prime }%
\tilde{b}_{1}^{\prime }\,,  \label{abe61a} \\
&&a_{-1}=\rho _{R}^{-\left( l+\mu \right) /2}c^{\prime }\tilde{a}%
_{-1}^{\prime },{}\;b_{-1}=\rho _{R}^{\left( l+\mu \right) /2}c^{\prime }%
\tilde{b}_{-1}^{\prime }\,,  \label{abe61b}
\end{eqnarray}
where the non-vanishing coefficients $\tilde{a}$, $\tilde{b}$, $\tilde{a}%
^{\prime }$, $\tilde{b}^{\prime }$ are independent from $\rho _{R}$ and the
coefficients $c$ and $c^{\prime }$ are normalization factors which depend on 
$\rho _{R}$.

Calculating the normalization factors, in the limit $R\rightarrow 0$, we
obtain 
\begin{eqnarray*}
a_{1} &=&\mathrm{const}\neq 0,\;b_{1}=0,\;a_{-1}=\;b_{-1}=0,\;l\geq 1\,, \\
a_{1} &=&0,\;b_{1}=\mathrm{const}\neq 0,\;a_{-1}=\;b_{-1}=0,\;l\leq 0
\end{eqnarray*}
for $l-l_{0}\leq 0$ and 
\begin{eqnarray*}
a_{1} &=&b_{1}=0,\;a_{-1}=\mathrm{const}\neq 0,\;b_{-1}=0,\;l\geq 0\,, \\
a_{1} &=&b_{1}=0,\;a_{-1}=0,\;b_{-1}=\mathrm{const}\neq 0,\;l\leq -1
\end{eqnarray*}
for $l-l_{0}>0$. For $l=0$, the value of the coefficients is defined by 
\textrm{sgn}$\Phi $. We can verify that the condition (\ref{x.2}) is
satisfied.

We thus obtain that for any sign of $B$, the solutions are expressed via
Laguerre polynomials (\ref{abe31}). In particular, for $l=0$, we find that\
the solutions $\psi _{\omega ,0}^{out}\left( r\right) $ coincide with either 
$\psi _{m}^{I}\left( r\right) $ or $\psi _{m}^{II}\left( r\right) $ in
accordance with $\mathrm{sgn}\left( \Phi \right) $, 
\begin{equation}
\psi _{\omega ,0}^{out}\left( r\right) =\left\{ 
\begin{array}{c}
\psi _{m}^{I}\left( r\right) ,\;\mathrm{sgn}\left( \Phi \right) =+1 \\ 
\psi _{m}^{II}\left( r\right) ,\;\mathrm{sgn}\left( \Phi \right) =-1
\end{array}
\right. \,.  \label{abe62}
\end{equation}

In Sec. 3, we have found the relation between the extension parameter values
and the types of solutions in the critical subspace $l=0$ (\ref{abe50a}), (%
\ref{abe50b}). We are now in a position to refine this relation. Namely, if
we introduce the AB field as the field of the zero-radius limit for the
finite-radius solenoid , then the extension parameter $\Theta $ is fixed to
be $\Theta =-\mathrm{sgn}\left( \Phi \right) \pi /2$. In addition, this way
of introducing the AB field implies no additional interaction in the
solenoid core.

To solve the problem in $3+1$ dimensions, we use the results in $2+1$
dimensions presented above. In the limit $R\rightarrow 0$, the\ solutions in
the critical subspace are 
\begin{equation}
\Psi _{s}^{out}\left( x_{\bot }\right) =N\left( 
\begin{array}{c}
\left[ 1+\left( p^{3}+s\widetilde{M}\right) /M\right] g_{0}\left( \varphi
\right) \psi _{\omega ,l}^{out}\left( r\right) \\ 
\left[ -1+\left( p^{3}+s\widetilde{M}\right) /M\right] g_{0}\left( \varphi
\right) \psi _{\omega ,l}^{out}\left( r\right)
\end{array}
\right) \,,  \label{r26}
\end{equation}
where the functions $g_{0}\left( \varphi \right) $, $\psi _{\omega
,l}^{out}\left( r\right) $ are defined in (\ref{abe11}) and\ (\ref{abe62}), (%
\ref{abe31}), respectively. The values of the extension parameters in $3+1$
dimensions are specified to be 
\begin{equation}
\Theta _{+1}=\Theta _{-1}=-\frac{\pi }{2}\mathrm{sgn\,}\Phi \,.  \label{r27}
\end{equation}

The interpretation of other possible $\Theta $'s via the limiting process
for other regularized potentials is not reached so far.

\subsection{Spinless particle case}

For completeness, let us consider the regularization problem in the spinless
particle case. For this case the Klein-Gordon equation with the
magnetic-solenoid field is, in fact, reduced to the eigenvalue problem for
the nonrelativistic two-dimensional Hamiltonian. Therefore, the 
self-adjoint extension problem as well as the solenoid regularization
problem is similar for the relativistic and nonrelativistic case.

From the classical paper \cite{AB59} on, the AB effect in the spinless case
was always associated with the radial functions regular at $r=0$. However,
the linkage between these kind of boundary conditions and a regularizing
solenoid is an interesting important problem. The aim of this subsection is
just to study this problem.

One ought to say  self-adjoint extensions of the nonrelativistic spinless
Hamiltonian with the AB field were studied in several articles \cite
{T74,MT91,GHK91,GMS96}. The case of the magnetic-solenoid field was
considered in \cite{ESV02} where the most general four-parameter family of
admissible boundary conditions was obtained. The solenoid regularization
with some particular distributions of the magnetic field inside the solenoid
was studied in \cite{T74,H90}. As we know, the regularization problem with
the arbitrary field inside the solenoid was not solved.

Unlike the Dirac equation, the Klein-Gordon equation was not solved
explicitly for the arbitrary field inside the regularizing solenoid.
However, one can obtain some properties of the corresponding solutions
without the explicit solving the Klein-Gordon equation. Thus, we can
demonstrate that for the arbitrary field inside the solenoid, the lifting of
the regularization ($R\rightarrow 0$) corresponds to regular radial
functions only.

To do that we use the above formulated in the Dirac case method. Solutions
of the Klein-Gordon equation with a given energy $\varepsilon $ and angular
momentum $l-l_{0}$ have the form 
\begin{equation*}
\phi \left( x\right) =Ne\,^{i\varepsilon x^{0}}e^{i\left( l-l_{0}\right)
\varphi }\phi \left( r\right) \,.
\end{equation*}
We find the radial function $\phi \left( r\right) $ sewing solutions inside
and outside the solenoid.

The outside solutions have the form \cite{BGT01}, 
\begin{equation}
\phi ^{out}\left( r\right) =\psi _{\lambda ,\alpha /2}\left( \rho \right)
=\rho ^{-1/2}W_{\lambda ,\alpha /2}\left( \rho \right) ,\;2\lambda =\omega
/\gamma -\xi \alpha ,\;\alpha =l+\mu \,,  \label{ss1}
\end{equation}
where, as before, $\omega =\varepsilon ^{2}-M^{2}$. The critical subspace is
defined by $l=0,-1$. The inside solutions satisfy the equation (\ref{abe58})
where one has to set $\sigma =0$. The approximation $\rho _{R}\ll 1$ and $%
\omega R^{2}\ll 1$\ can be applied in this case as well. Therefore, we
arrive to the following equation for the inside solutions, 
\begin{equation}
\left[ \frac{d^{2}}{dx^{2}}+\frac{1}{x}\frac{d}{dx}-\frac{1}{x^{2}}\left(
\eta +\vartheta \left( x\right) \right) ^{2}\right] \phi ^{in}\left(
x\right) =0\,,  \label{s3}
\end{equation}
where $0\leq x\leq 1$. We are looking for solutions $\phi ^{in}\left(
x\right) $ that are regular at $x=0$. Applying the sewing conditions (\ref
{cond-cont}), we obtain 
\begin{equation}
\alpha \phi _{in}\left( 1\right) +\partial _{x}\phi ^{in}\left( 1\right)
=2\alpha a\rho _{R}^{\alpha /2}\,,\;\alpha \phi _{in}\left( 1\right)
-\partial _{x}\phi ^{in}\left( 1\right) =2\alpha b\rho _{R}^{-\alpha /2}
\label{s1}
\end{equation}
in the lower order in $\rho _{R}$.

For our purposes it is important to demonstrate that the following condition
holds true 
\begin{equation}
\left| \alpha \right| \phi ^{in}\left( 1\right) +\partial _{x}\phi
^{in}\left( 1\right) \neq 0\,.  \label{s2}
\end{equation}
To this end, let us find solutions of equation (\ref{s3}) that are regular
at $x=0$. The function $\vartheta \left( x\right) $\ is analytic on the
interval $0\leq x\leq 1$ as obeying the conditions listed in Sec. 4.1. Then (%
\ref{s3}) is the homogeneous ordinary differential equation with the regular
singular point $x=0$. It is known from the general theory \cite{Korn} that
there exist solutions of (\ref{s3}) that can be represented in the form 
\begin{equation}
\phi ^{in}\left( x\right) =x^{\left| \eta \right| }\sum_{k=0}^{\infty
}a_{k}x^{k}\,,  \label{s4}
\end{equation}
where $a_{0}$ is an arbitrary constant and the other coefficients are
defined via recurrent relations. The series in (\ref{s4}) is absolutely and
uniformly convergent for $0\leq x\leq 1$. Let us suppose now for simplicity
that $f\left( x\right) $ does not change its sign for $0\leq x\leq 1$. Then
we represent $\phi ^{in}\left( x\right) $ in the form 
\begin{equation}
\phi _{in}\left( x\right) =\left\{ 
\begin{array}{c}
x^{\left| \eta \right| }e^{-\left| b\left( x\right) \right| }\varphi
^{\left( -\right) }\left( x\right) ,\;\mathrm{sgn}\left( f\right) \eta \leq 0
\\ 
x^{-\left| \eta \right| }e^{-\left| b\left( x\right) \right| }\varphi
^{\left( +\right) }\left( x\right) ,\;\mathrm{sgn}\left( f\right) \eta >0
\end{array}
\right. ,  \label{s7}
\end{equation}
where the functions $\varphi ^{\left( \pm \right) }\left( x\right) $ obey
the equation 
\begin{equation}
H^{\left( \pm \right) }\varphi ^{\left( \pm \right) }=0\,,\;H^{\left( \pm
\right) }=H_{0}^{\left( \pm \right) }-\left| f\left( x\right) \right|
\,,\;H_{0}^{\left( \pm \right) }=\frac{d^{2}}{dx^{2}}+\left( 1\mp 2\left|
\eta \right| -2\left| \vartheta \left( x\right) \right| \right) \frac{1}{x}%
\frac{d}{dx}  \label{s8}
\end{equation}
and the conditions 
\begin{equation}
\varphi ^{\left( -\right) }\left( 0\right) =1,\;\left. x^{-2\left| \eta
\right| }\varphi ^{\left( +\right) }\left( x\right) \right| _{x=0}=1\,.
\label{s10}
\end{equation}
It is convenient to introduce into the consideration the retarded propagator 
$G_{ret}^{\left( \pm \right) }\left( x,y\right) $ obeying the equation 
\begin{equation*}
H_{0}^{\left( \pm \right) }G_{ret}^{\left( \pm \right) }\left( x,y\right)
=\delta \left( x-y\right) \,.
\end{equation*}
It can be represented as $G_{ret}^{\left( \pm \right) }\left( x,y\right)
=\theta \left( x-y\right) G^{\left( \pm \right) }\left( x,y\right) $ where
the functions $G^{\left( \pm \right) }$ obey the conditions 
\begin{equation*}
G^{\left( \pm \right) }\left( x,x\right) =0,\;\left. \frac{\partial }{%
\partial x}G^{\left( \pm \right) }\left( x,y\right) \right| _{x=y}=1\,,
\end{equation*}
and can be found explicitly in the case under consideration, 
\begin{equation}
G^{\left( \pm \right) }\left( x,y\right) =\int_{y}^{x}\left( \frac{%
\widetilde{x}}{y}\right) ^{\pm 2\left| \eta \right| -1}\exp \left\{ 2\left|
b\left( \widetilde{x}\right) \right| -2\left| b\left( y\right) \right|
\right\} d\widetilde{x}\,.  \label{s11}
\end{equation}
The differential equation (\ref{s8}) with the boundary conditions (\ref{s10}%
) is equivalent to the following integral equation, 
\begin{equation}
\varphi ^{\left( \pm \right) }\left( x\right) =\varphi _{0}^{\left( \pm
\right) }\left( x\right) +\int_{0}^{x}G^{\left( \pm \right) }\left(
x,y\right) \left| f\left( y\right) \right| \varphi ^{\left( \pm \right)
}\left( y\right) dy\,,  \label{s12}
\end{equation}
where 
\begin{equation*}
\varphi _{0}^{\left( -\right) }\left( x\right) =1,\;\varphi _{0}^{\left(
+\right) }\left( x\right) =\int_{0}^{x}\widetilde{x}^{2\left| \eta \right|
-1}e^{2\left| b\left( \widetilde{x}\right) \right| }d\widetilde{x}\,.
\end{equation*}
The solutions of the equation (\ref{s12}) can be found by iterations, 
\begin{eqnarray}
\varphi ^{\left( \pm \right) }\left( x\right) &=&\sum_{k=0}^{\infty
}Y_{k}^{\left( \pm \right) }\left( x\right) \,,\;  \notag \\
Y_{k}^{\left( \pm \right) }\left( x\right) &=&\int_{0}^{x}G^{\left( \pm
\right) }\left( x,y\right) \left| f\left( y\right) \right| Y_{k-1}^{\left(
\pm \right) }\left( y\right) dy,\;Y_{0}^{\left( \pm \right) }\left( x\right)
=\varphi _{0}^{\left( \pm \right) }\left( x\right) \,.  \label{s13}
\end{eqnarray}
Every member of the series (\ref{s13}) is positive. The series converges
uniformly. Then, it implies $\varphi ^{\left( \pm \right) }\left( x\right)
>0 $ and $\partial _{x}\varphi ^{\left( \pm \right) }\left( x\right) >0$ for 
$x>0$. Therefore, 
\begin{equation}
\phi ^{in}\left( 1\right) >0,\;\partial _{x}\phi ^{in}\left( 1\right) =-%
\mathrm{sgn}\left( f\right) \alpha \phi ^{in}\left( 1\right) +e^{-\left|
b\left( 1\right) \right| }\partial _{x}\varphi ^{\left( \pm \right) }\left(
1\right) \,.  \label{s14}
\end{equation}
If \textrm{sgn}$\left( f\right) \alpha <0,$ then $\partial _{x}\phi
^{in}\left( 1\right) >0$\ and the condition (\ref{s2}) is satisfied. If 
\textrm{sgn}$\left( f\right) \alpha >0$, then it follows from (\ref{s14})
that 
\begin{equation*}
\left| \alpha \right| \phi ^{in}\left( 1\right) +\partial _{x}\phi
^{in}\left( 1\right) >0\,,
\end{equation*}
and the condition (\ref{s2}) is\ satisfied as well. The obtained result can
be extended to the function $f\left( x\right) $ alternating in sign. In this
case the interval $\left[ 0,1\right] $ can be divided into subintervals on
which $f\left( x\right) $ does not change its sign. The solutions $\phi
^{in}\left( x\right) $ can be found successively on each such a subinterval
beginning from the point $x=0$.

Applying the normalization condition for the sewed solutions, 
\begin{equation*}
\int_{0}^{\rho _{R}}\left| \phi ^{in}\right| ^{2}d\rho +\int_{\rho
_{R}}^{\infty }\left| \phi ^{out}\right| ^{2}d\rho =1\,,
\end{equation*}
we get 
\begin{equation}
b=0\;\text{if}\;\alpha >0;\;a=0\;\text{if}\;\alpha <0  \label{ss5}
\end{equation}
in the limit $R\rightarrow 0$. Thus, the only regular at $r=0$ solutions
remain in the limit $R\rightarrow 0$. The conditions (\ref{ss5}) also define
the spectrum. Finally, it follows from (\ref{abe59}) that in the limit $%
R\rightarrow 0$ the radial functions and the related spectrum has the form 
\begin{equation}
\phi _{m,l}\left( r\right) =I_{m+\left| l+\mu \right| ,m}\left( \rho \right)
\,,\;\omega =\gamma \left( 2m+\left| l+\mu \right| +\xi \left( l+\mu \right)
+1\right) \,.  \label{g50}
\end{equation}

\section{Reduced  self-adjoint extension method}

We here show that the general  self-adjoint extension method can be
significantly reduced for the radial Hamiltonian $h$ given by Eqs. (\ref
{abe12}), (\ref{abe13}).

We recall that the problem is to define the formal matrix differential
operator (\ref{abe12}), (\ref{abe13}) as a self-adjoint operator in the
Hilbert space of two-spinors $\chi \left( r\right) $ square-integrable on
the half-line $R_{+}=\{r\geq 0\}$ with the measure $rdr.$ It is convenient
to pass to the standard measure $dr$ on $R_{+}$ with the substitution $\chi
\left( r\right) =r^{-1/2}\Psi \left( r\right) \,.$ The radial Hamiltonian
then becomes (we do not change the notation for the transformed Hamiltonian $%
h$) 
\begin{equation}
h=\Pi +\sigma ^{3}M\,,\;\;\Pi =\left( 
\begin{array}{cc}
0 & a_{\nu ,B} \\ 
a_{-\nu ,-B} & 0
\end{array}
\right) \,,  \label{85}
\end{equation}
where $\nu =\mu +l-1/2$ and the operator $a_{\nu ,B}$: $L_{2}\rightarrow
L_{2}$ is 
\begin{equation}
a_{\nu ,B}=-i\left( \partial _{r}+\frac{\nu }{r}+\frac{eBr}{2}\right)
=-i\sigma _{-\nu ,-B}\partial _{r}\sigma _{\nu ,B}  \label{86}
\end{equation}
with 
\begin{equation*}
\sigma _{\nu ,B}=r^{\nu }\exp \left( \frac{eBr^{2}}{4}\right) \,,\;\sigma
_{-\nu ,-B}\sigma _{\nu ,B}=1\,.
\end{equation*}
This $h$ (\ref{85}) must be defined as a self-adjoint operator in the
Hilbert space 
\begin{equation*}
\mathcal{H}=L_{2}\left( R_{+}\right) \oplus L_{2}\left( R_{+}\right) 
\end{equation*}
of two-spinors 
\begin{equation*}
\Psi \left( r\right) =\left( 
\begin{array}{c}
\psi ^{1}\left( r\right)  \\ 
\psi ^{2}\left( r\right) 
\end{array}
\right) 
\end{equation*}
square integrable on $R_{+\,}$.

We first note that $\sigma ^{3}M$ is a bounded self-adjoint operator in $%
\mathcal{H}.$ Therefore, the problem is equivalent to the problem of
defining the apparently self-adjoint\footnote{%
Formally, $a_{\nu ,B}^{\dagger }=a_{-\nu ,-B}\,,$ therefore, again formally, 
\begin{equation*}
\Pi ^{\dagger }=\left( 
\begin{array}{cc}
0 & a_{-\nu ,-B}^{\dagger } \\ 
a_{\nu ,B}^{\dagger } & 0
\end{array}
\right) =\left( 
\begin{array}{cc}
0 & a_{\nu ,B} \\ 
a_{-\nu ,-B} & 0
\end{array}
\right) =\Pi \,.
\end{equation*}
} operator $\Pi $ (\ref{85}) as a really self-adjoint operator in $\mathcal{H%
}:$ we must ensure the equality $\Pi =\Pi ^{\dagger }$ by the proper choice
of the domain 
\begin{equation*}
\mathcal{D}(\Pi )=\mathcal{D}(a_{-\nu ,-B})\oplus \mathcal{D}(a_{\nu
,B})\subset \mathcal{H}\,,
\end{equation*}
for any $\nu $ and $B$. The problem is thus, reduced to the problem of
properly defining $a_{\nu ,B}$ (\ref{86}) as an operator in $L_{2}\left(
R_{+}\right) ,$ i.e., defining $\mathcal{D}(a_{\nu ,B})\subset L_{2}\left(
R_{+}\right) ,$ together with $a_{-\nu ,-B}\,$.

We canonically start with defining $\Pi $ as a symmetric operator in $%
\mathcal{H}$ by taking the initial domain to be $\mathcal{D}(\Pi )=D\oplus D$%
, such that the both $a_{\nu ,B}$ and $a_{-\nu ,-B}$ are initially defined
on $D\subset L_{2}\left( R_{+}\right) ,$ where $D$ is the linear space of $%
C^{\infty }$-functions with compact support. We note that $\Pi $ is densely
defined because $\overline{D}\subset L_{2}\left( R_{+}\right) $ and
symmetric, $\Pi \subseteq \Pi ^{\dagger }$,\ which is simply verified by
integration by parts.

The next step is the evaluation of its adjoint \thinspace $\Pi ^{\dagger }.$
It is evident that 
\begin{equation}
\,\Pi ^{\dagger }=\left( 
\begin{array}{cc}
0 & a_{-\nu ,-B}^{\dagger } \\ 
a_{\nu ,B}^{\dagger } & 0
\end{array}
\right) \,,\;\;\mathcal{D}(\,\Pi ^{\dagger })=\mathcal{D}(a_{\nu
,B}^{\dagger })\oplus \mathcal{D}(a_{-\nu ,-B}^{\dagger })\,,  \label{87}
\end{equation}
for any $\nu $ and $B.$ Because $a_{\nu ,B}$ is a simple differential
operator in $L_{2}\left( R_{+}\right) ,$ its adjoint $a_{\nu ,B}^{\dagger }$
is evaluated simply (by the standard method for differential operators in $%
L_{2}\left( R_{+}\right) $): its form is 
\begin{equation}
a_{\nu ,B}^{\dagger }=-i\sigma _{\nu ,B}\,\partial _{r}\sigma _{-\nu ,-B}
\label{88}
\end{equation}
(it coincides with $a_{-\nu ,-B}$ (\ref{86})), and its domain is 
\begin{equation}
\mathcal{D}(a_{\nu ,B}^{\dagger })=\{_{\ast }\psi _{\nu ,B}\left( r\right)
\}\,,  \label{89}
\end{equation}
where $_{\ast }\psi _{\nu ,B}\left( r\right) $ are absolutely continuous
(inside $R_{+}$) square integrable functions allowing the representation 
\begin{equation}
_{\ast }\psi _{\nu ,B}\left( r\right) =i\sigma _{\nu ,B}\left( r\right) 
\left[ \int_{r_{0}}^{r}d\xi \,\sigma _{-\nu ,-B}\left( \xi \right) _{\ast
}\phi _{\nu ,B}\left( \xi \right) +c_{\nu ,B}\right] \,.  \label{90}
\end{equation}
Here, $_{\ast }\phi _{\nu ,B}=a_{\nu ,B}^{\dagger }\,_{\ast }\psi _{\nu ,B},$
the image of $a_{\nu ,B}^{\dagger }$ on $_{\ast }\psi _{\nu ,B}$ , is square
integrable, $_{\ast }\phi _{\nu ,B}\in L_{2}\left( R_{+}\right) ;\;c_{\nu
,B} $ is a constant restricted by the requirement that $_{\ast }\psi _{\nu
,B}\in L_{2}\left( R_{+}\right) ;\;r_{0}\geq 0$ is chosen appropriately
depending on the value of $\nu $ and the sign of $B.$ Of course, the same is
true for $a_{-\nu ,-B}^{+}:$ we must simply make the substitutions $\nu
\rightleftarrows -\nu ,\;B\rightleftarrows -B$ in (\ref{88}-\ref{90}).

At this point, we depart from the general procedure used in Sec. 3: then
finding the deficiency subspaces and etc. Instead, we determine the
''asymmetry'' of \thinspace $\Pi ^{\dagger }$ evaluating the difference 
\begin{equation}
\Delta =\left( _{\ast }\Psi ^{\prime },\,\Pi ^{\dagger }{}_{\ast }\Psi
\right) -\left( \,\Pi ^{\dagger }{}_{\ast }\Psi ^{\prime },_{\ast }\Psi
\right)  \label{91}
\end{equation}
for any $_{\ast }\Psi $ and${}_{\ast }\Psi ^{\prime }$ belonging to $%
\mathcal{D}($\thinspace $\Pi ^{\dagger }).$ Using the form of $_{\ast }\Psi
\in \mathcal{D}($\thinspace $\Pi ^{\dagger }),$ see (\ref{87}), (\ref{89}), (%
\ref{90}), 
\begin{equation}
_{\ast }\Psi \left( r\right) =\left( 
\begin{array}{c}
_{\ast }\psi _{\nu ,B}^{1}\left( r\right) \\ 
_{\ast }\psi _{-\nu ,-B}^{2}\left( r\right)
\end{array}
\right)  \label{92}
\end{equation}
and the form (\ref{87}) of \thinspace $\Pi ^{\dagger },$ we obtain (we here
omit the subscripts $\ast $ and $B$ as irrelevant, which becomes clear below)

\begin{eqnarray*}
&&\Delta =\int_{0}^{\infty }dr\left[ \overline{\psi _{\nu }^{\prime 1}\left(
r\right) \,}a_{-\nu }^{\dagger }\psi _{-\nu }^{2}\left( r\right) +\overline{%
\psi _{-\nu }^{\prime 2}\left( r\right) \,}a_{\nu }^{\dagger }\psi _{\nu
}^{1}\left( r\right) \right] \\
&&-\int_{0}^{\infty }dr\left[ \overline{a_{-\nu }^{\dagger }\psi _{-\nu
}^{\prime 2}\left( r\right) }\,\psi _{\nu }^{1}\left( r\right) +\overline{%
a_{\nu }^{\dagger }\psi _{\nu }^{\prime 1}\left( r\right) \,}\psi _{-\nu
}^{2}\left( r\right) \right] \,.
\end{eqnarray*}
Integrating by parts in the first integral, we find 
\begin{equation*}
\Delta =-i\left. \left[ \overline{\psi _{\nu }^{\prime 1}\left( r\right) }%
\psi _{-\nu }^{2}\left( r\right) +\overline{\psi _{-\nu }^{\prime 2}\left(
r\right) }\psi _{\nu }^{1}\left( r\right) \right] \right| _{0}^{\infty }\,,
\end{equation*}
i.e., $\Delta $ is determined by the asymptotic behavior of $_{\ast }\Psi $ (%
\ref{92}) at the boundaries, as $r\rightarrow \infty $ and $r\rightarrow 0.$
The asymptotic behavior for $_{\ast }\psi _{\nu ,B}\left( r\right) $ and $%
_{\ast }\psi _{-\nu ,-B}\left( r\right) $ can be estimated using the
representation (\ref{90}) with the appropriate choice of $r_{0}$ and
estimating the integral term in (\ref{90}) via the Cauchy-Bounjakowsky
inequality.

For example, we estimate the behavior of $_{\ast }\psi _{\nu ,B}\left(
r\right) $ at infinity, as $r\rightarrow \infty ,$ in the case $B>0.$ In
this case, it is convenient to take $r_{0}=\infty ,$ such that 
\begin{equation}
_{\ast }\psi _{\nu ,B}\left( r\right) =i\sigma _{\nu ,B}\left( r\right) 
\left[ -\int_{r}^{\infty }d\xi \,\sigma _{-\nu ,-B}\left( \xi \right) _{\ast
}\phi _{\nu ,B}\left( \xi \right) +c_{\nu ,B}\right] \,.  \label{93}
\end{equation}
The Cauchy-Bounjakowsky inequality yields 
\begin{eqnarray}
&&\left| \int_{r}^{\infty }d\xi \xi ^{-\nu }\exp \left( -\frac{eB\xi ^{2}}{4}%
\right) \right| \leq \left[ \int_{r}^{\infty }d\xi \,\xi ^{-2\nu }\exp
\left( -\frac{eB\xi ^{2}}{4}\right) \,_{\ast }\phi _{\nu ,B}\left( \xi
\right) \right] ^{1/2}  \notag \\
&&\,\times \left[ \int_{r}^{\infty }d\xi \left| _{\ast }\phi _{\nu ,B}\left(
\xi \right) \right| ^{2}\right] ^{1/2}=\frac{r^{-\nu -1/2}}{eB}\exp \left( -%
\frac{eBr^{2}}{4}\right) \left( 1+O\left( \frac{1}{r}\right) \right)
\epsilon \left( r\right) \,,  \label{94}
\end{eqnarray}
where $\epsilon \left( r\right) \rightarrow 0$ as $r\rightarrow \infty .$ It
follows from (\ref{94}) and the condition $_{\ast }\psi _{\nu ,B}\in
L_{2}\left( R_{+}\right) $ that $c_{\nu ,B}$ in (\ref{93}) must be zero and,
consequently, $_{\ast }\psi _{\nu ,B}\left( r\right) $ vanishes as $%
r\rightarrow \infty ,$ faster than $r^{-1/2},$ independently from $\nu .$
The same is true for $B<0,$ which is established via the Cauchy-Bounjakowsky
inequality with $r_{0}<\infty .$

The asymptotic estimates for $_{\ast }\psi _{\nu ,B}\left( r\right) $ as $%
r\rightarrow 0$ depend on the value of $\nu .$ The result is: 
\begin{eqnarray*}
\nu >\frac{1}{2} &:&\;\left| _{\ast }\psi _{\nu ,B}\left( r\right) \right|
<c_{\nu }r^{1/2}\,, \\
\nu =\frac{1}{2} &:&\;\left| _{\ast }\psi _{1/2,B}\left( r\right) \right|
<c_{1/2}\,r^{1/2}\ln \frac{r}{r_{0}}\,, \\
\nu \leq -\frac{1}{2} &:&\;_{\ast }\psi _{\nu ,B}\left( r\right) =o\left(
r^{1/2}\right) \,, \\
-\frac{1}{2}<\nu <\frac{1}{2} &:&\;_{\ast }\psi _{\nu ,B}\left( r\right)
=c_{\nu ,B}\,\left( rM\right) ^{\nu }+o\left( r^{1/2}\right) \,,
\end{eqnarray*}
independently from $B;$ the factor $M$ is introduced by the dimensional
reasons.

We conclude that for $\left| \nu \right| \geq 1/2\,,$ the two-spinors $%
_{\ast }\Psi \left( r\right) $ (\ref{92}) vanish both at infinity and at the
origin, whereas for $\left| \nu \right| <1/2\,,\;$the two-spinors $_{\ast
}\Psi \left( r\right) $ vanish at infinity, but are generally singular as $%
r\rightarrow 0:$%
\begin{equation}
_{\ast }\Psi \left( r\right) =\left( 
\begin{array}{c}
c_{\nu ,B}^{1}\,\left( rM\right) ^{\nu } \\ 
c_{-\nu ,-B}^{2}\,\left( rM\right) ^{-\nu }
\end{array}
\right) +o\left( r^{1/2}\right) \,,  \label{95}
\end{equation}
where $c^{1,2}$ are arbitrary constants. With this estimates in hand, we
find 
\begin{equation}
\Delta =\left\{ 
\begin{array}{c}
0\,,\;\left| \nu \right| \geq 1/2 \\ 
i\left[ c_{\nu }^{\prime 1}c_{-\nu }^{2}+c_{-\nu }^{\prime 2}c_{\nu }^{1}%
\right] \,,\;\left| \nu \right| <1/2\,
\end{array}
\right. ,  \label{96}
\end{equation}
independently of $B$ (we therefore omit the subscripts $\ast $ and $B$ as
irrelevant).

We thus find that if $\left| \nu \right| \geq 1/2\,,$ i.e., if $l\neq 0$ or $%
l=0$ and $\mu =0$, the operator $\Pi ^{\dagger }$ is symmetric, $\Pi
^{\dagger }\subseteq \left( \Pi ^{\dagger }\right) ^{\dagger }$, and
therefore  self-adjoint, $\Pi ^{\dagger }=\left( \Pi ^{\dagger }\right)
^{\dagger }$: it is sufficient to take the inverse inclusion $\left( \Pi
^{\dagger }\right) ^{\dagger }\subseteq \Pi ^{\dagger }$ into account, the
standard relation for any symmetric operator (which is the consequence of
the general relations $A\subseteq B\rightarrow B^{\dagger }\subseteq
A^{\dagger }$ for any densely defined operators). Thus, $\Pi ^{\dagger }$\
is a  self-adjoint extension of $\Pi $.

This implies that $\Pi $ is essentially  self-adjoint, i.e., its closure $%
\overline{\Pi }$ is  self-adjoint, $\overline{\Pi }=\overline{\Pi }^{\dagger
}$, in addition $\overline{\Pi }=\Pi ^{\dagger }$ and is a unique 
self-adjoint extension of $\Pi $. We recall the standard arguments. As is
well known, $\overline{\Pi }=\left( \Pi ^{\dagger }\right) ^{\dagger }$ and $%
\overline{\Pi }^{\dagger }=\Pi ^{\dagger }$. Using the established 
self-adjointness of $\Pi ^{\dagger }$, $\Pi ^{\dagger }=\left( \Pi ^{\dagger
}\right) ^{\dagger }$, we obtain the chain of equalities $\overline{\Pi }%
=\left( \Pi ^{\dagger }\right) ^{\dagger }=\Pi ^{\dagger }=\overline{\Pi }%
^{\dagger }$, which proves the first two assertions. There are no other 
self-adjoint extensions because any such extension $\Pi ^{ext}=\left( \Pi
^{ext}\right) ^{\dagger }$ must satisfy the relation $\overline{\Pi }%
\subseteq \Pi ^{ext}\subseteq \Pi ^{\dagger }$, $\Pi ^{ext}$\ must be the
extension of $\overline{\Pi }$\ and the restriction of $\Pi ^{\dagger }$,
but because $\overline{\Pi }=\Pi ^{\dagger }$, there are ''no place '' for
another  self-adjoint extension.

The situation is nontrivial in the case $\left| \nu \right| <1/2$, i.e. if $%
l=0$ and $\mu \neq 0$, where the operator $\Pi ^{\dagger }$\ is
nonsymmetric, which implies that $\overline{\Pi }$ is only symmetric: $%
\overline{\Pi }=\left( \Pi ^{\dagger }\right) ^{\dagger }\subset \Pi
^{\dagger }=\overline{\Pi }^{\dagger }$ is a strict inclusion. This relation
allows finding $\overline{\Pi }$. The inclusion $\overline{\Pi }\subset \Pi
^{\dagger }$\ implies that $\mathcal{D}\left( \overline{\Pi }\right) \subset 
\mathcal{D}\left( \Pi ^{\dagger }\right) $, i. e. $\Psi ^{\prime }\in $ $%
\mathcal{D}\left( \overline{\Pi }\right) $ has the representations like (\ref
{90}) and (\ref{95}), 
\begin{equation}
\Psi ^{\prime }\left( r\right) =\left( 
\begin{array}{c}
c^{\prime 1}\left( rM\right) ^{\nu } \\ 
c^{\prime 2}\left( rM\right) ^{-\nu }
\end{array}
\right) +o\left( r^{1/2}\right) ,  \label{97}
\end{equation}
and vanish at infinity. Then the equality $\overline{\Pi }=\left( \Pi
^{\dagger }\right) ^{\dagger }\subset \Pi ^{\dagger }$\ implies, by the
definition of $\left( \Pi ^{\dagger }\right) ^{\dagger }$, that $\Psi
^{\prime }\in \mathcal{D}\left( \overline{\Pi }\right) $ iff the difference $%
\Delta $ (\ref{91}) vanishes for $\Psi ^{\prime }\in \mathcal{D}\left( 
\overline{\Pi }\right) \subset \mathcal{D}\left( \Pi ^{\dagger }\right) $
and any $\Psi \in \mathcal{D}\left( \Pi ^{\dagger }\right) $. According to (%
\ref{96}), this gives 
\begin{equation*}
\overline{c_{\nu }^{\prime 1}}c_{-\nu }^{2}+\overline{c_{-\nu }^{\prime 2}}%
c_{\nu }^{1}=0,\;\forall c_{\nu }^{1},\;c_{-\nu }^{2}\,,
\end{equation*}
whence $c_{\nu }^{\prime 1}=c_{-\nu }^{\prime 2}=0$. $\overline{\Pi }$\ is
thus defined as the restriction of $\Pi ^{\dagger }$ to the domain $\mathcal{%
D}\left( \overline{\Pi }\right) $\ of the functions belonging to $\mathcal{D}%
\left( \Pi ^{\dagger }\right) $, but vanishing at the origin.

We must seek the  self-adjoint extensions $\Pi ^{ext}=\left( \Pi
^{ext}\right) ^{\dagger }$ of $\Pi $. These, as was said just before, ''must
lie between'' $\overline{\Pi }$ and $\Pi ^{\dagger }$, $\overline{\Pi }%
\subset \Pi ^{ext}=\left( \Pi ^{ext}\right) ^{\dagger }\subset \Pi ^{\dagger
}$. We start with finding the nontrivial (not coinciding with $\overline{\Pi 
}$) symmetric extensions $\Pi ^{sym}$ of $\Pi $, $\overline{\Pi }\subset \Pi
^{sym}\subseteq \left( \Pi ^{sym}\right) ^{\dagger }\subseteq \Pi ^{\dagger }
$ and then show that these $\Pi ^{sym}$\ solve the problem. The last
inclusions allows repeating for $\Pi ^{sym}$\ the previous arguments for $%
\overline{\Pi }$\ except that now the difference $\Delta $ (\ref{91}) must
vanish for any $\Psi $, $\Psi ^{\prime }\in \mathcal{D}\left( \Pi
^{sym}\right) \subset \mathcal{D}\left( \Pi ^{\dagger }\right) $. According
to (\ref{96}) this gives the equation for $c_{\nu }^{1}$,$\;c_{-\nu }^{2}$
in $\Psi $ (\ref{95}) and $c_{\nu }^{\prime 1}$, $c_{-\nu }^{\prime 2}$\ in $%
\Psi ^{\prime }$ (\ref{97}), 
\begin{equation}
\overline{c_{\nu }^{\prime 1}}c_{-\nu }^{2}+\overline{c_{-\nu }^{\prime 2}}%
c_{\nu }^{1}=0,  \label{98}
\end{equation}
Of course, if $\Psi $, $\Psi ^{\prime }\in \mathcal{D}\left( \overline{\Pi }%
\right) \subset \mathcal{D}\left( \Pi ^{sym}\right) $, i.e. vanish at the
origin, Eq. (\ref{98}) holds. To satisfy (\ref{98}) it is sufficient that
only $\Psi $ or $\Psi ^{\prime }$\ belong to $\mathcal{D}\left( \overline{%
\Pi }\right) $. But $\mathcal{D}\left( \Pi ^{sym}\right) $ must contain
functions nonvanishing at the origin. Let $\Psi $ be such a function and,
for example, let $c_{-\nu }^{2}\neq 0$. For $\Psi ^{\prime }=\Psi $, Eq. (%
\ref{98}) becomes 
\begin{equation*}
\overline{c_{\nu }^{1}}c_{-\nu }^{2}+\overline{c_{-\nu }^{2}}c_{\nu }^{1}=0,
\end{equation*}
whence 
\begin{equation*}
\frac{c_{\nu }^{1}}{c_{-\nu }^{2}}=-\frac{\overline{c_{\nu }^{1}}}{\overline{%
c_{-\nu }^{2}}}=i\,\Lambda ,\;\,\Lambda \in \mathbb{R}.
\end{equation*}
Then for any $\Psi ^{\prime }\in \mathcal{D}\left( \Pi ^{sym}\right) $
nonvanishing at the origin and this fixed $\Psi $, Eq. (\ref{98}) becomes 
\begin{equation*}
\overline{c_{\nu }^{\prime 1}}+i\,\Lambda \overline{c_{-\nu }^{\prime 2}}=0,
\end{equation*}
whence $c_{-\nu }^{\prime 2}\neq 0$\ and $c_{\nu }^{\prime 1}/c_{-\nu
}^{\prime 2}=i$\thinspace $\Lambda $, with the same \thinspace $\Lambda $
for all $\Psi \in \mathcal{D}\left( \Pi ^{sym}\right) $ nonvanishing at the
origin.

The case $c_{-\nu }^{2}=0$, $c_{\nu }^{1}\neq 0$\ is considered similarly.
It is formally covered by the case \thinspace $\Lambda =\pm \infty $, where
\thinspace $\Lambda =+\infty $ and \thinspace $\Lambda =-\infty $ are
equivalent (both simply mean $c_{-\nu }^{2}=0$, $c_{\nu }^{1}\neq 0$). We
thus obtain the asymptotic boundary conditions 
\begin{equation}
\Psi \left( r\right) =c\left( 
\begin{array}{c}
i\,\Lambda \left( rM\right) ^{\nu } \\ 
\left( rM\right) ^{-\nu }
\end{array}
\right) +o\left( r^{1/2}\right)  \label{99}
\end{equation}
at the origin as $r\rightarrow 0$, that define a one-parameter family $%
\left\{ \Pi ^{\,\Lambda }\right\} $ of all nontrivial symmetric extensions
of $\Pi $.

It remains to show that these extensions $\Pi ^{\,\Lambda }$ are really 
self-adjoint, $\Pi ^{\,\Lambda }=\left( \Pi ^{\,\Lambda }\right) ^{\dagger }$%
. For this, it is necessary to evaluate $\left( \Pi ^{\Lambda }\right)
^{\dagger }$. Using the arguments similar to the previous ones, we conclude
that $\Psi ^{\prime }\in \mathcal{D}\left( \left( \Pi ^{\,\Lambda }\right)
^{\dagger }\right) $\ iff the difference $\Delta $ (\ref{91}) vanish for $%
\Psi ^{\prime }\in \mathcal{D}\left( \left( \Pi ^{\Lambda }\right) ^{\dagger
}\right) \subseteq \mathcal{D}\left( \Pi ^{\dagger }\right) $\ and any $\Psi
\in \mathcal{D}\left( \Pi ^{\Lambda }\right) \subset \mathcal{D}\left( \Pi
^{\dagger }\right) $, which yields 
\begin{equation*}
\overline{c_{\nu }^{\prime 1}}+i\,\Lambda \overline{c_{-\nu }^{\prime 2}}=0.
\end{equation*}
If $c_{\nu }^{\prime 1}$ and $c_{-\nu }^{\prime 2}$\ are not equal to zero,
this gives 
\begin{equation*}
\frac{c_{\nu }^{\prime 1}}{c_{-\nu }^{\prime 2}}=i\,\Lambda .
\end{equation*}
But this means that $\mathcal{D}\left( \left( \Pi ^{\Lambda }\right)
^{\dagger }\right) \subseteq \mathcal{D}\left( \Pi ^{\Lambda }\right) $,
i.e. $\left( \Pi ^{\,\Lambda }\right) ^{\dagger }\subseteq \Pi ^{\,\Lambda }$%
, the inclusion inverse to the initial $\Pi ^{\,\Lambda }\subseteq \left(
\Pi ^{\,\Lambda }\right) ^{\dagger }$, whence $\Pi ^{\,\Lambda }=\left( \Pi
^{\Lambda }\right) ^{\dagger }$, which proves the final statement.

We thus found the whole set of the  self-adjoint extensions of the initial
symmetric operator $\Pi $ (\ref{85}) with $\mathcal{D}\left( \Pi \right) =%
\mathcal{D}\oplus \mathcal{D}$. This is a one -parameter family $\left\{ \Pi
^{\Lambda }\right\} $, $-\infty \leq $\thinspace $\Lambda \leq \infty $,
whose each member $\Pi ^{\Lambda }=\left( \Pi ^{\Lambda }\right) ^{\dagger }$%
\ is defined by the asymptotic boundary condition (\ref{99}) at the origin
on $\Psi \in \mathcal{D}\left( \Pi ^{\Lambda }\right) \subset \mathcal{D}%
\left( \Pi ^{\dagger }\right) $. Because \thinspace $\Lambda =\pm \infty $
are equivalent, this family is homeomorphic to a circle $U\left( 1\right) $,
not an open line $\mathbb{R}$. The obtained  self-adjoint asymptotic
boundary conditions (\ref{99}) evidently coincide with the  self-adjoint
asymptotic boundary conditions (\ref{abe45}) (up to the common factor $%
M^{1/2}$) obtained by the general method. As a by product, we find that the
deficiency indices of $\Pi $ are $\left( 1,1\right) $\ in the case $\left|
\nu \right| <1/2$.

We conclude with the remark that those who is well aware of special
functions (to effectively determine the deficiency subspaces) may prefer the
general method (we note that, in fact, only the asymptotic behavior at the
origin of the corresponding functions is need). But in any case, the
evaluation of the adjoint operator, $\Pi ^{\dagger }$ or $h^{\dagger },$ is
imperative in order to correctly determine the domain of the  self-adjoint
extensions.

\section{Concluding remarks}

We have studied the solutions of the Dirac equation with the
magnetic-solenoid field in $2+1$ and $3+1$ dimensions in detail. In the
general case, there no simple relations between the solutions in $2+1$ and $%
3+1$ dimensions. However, we have demonstrated that the solutions in $3+1$
dimensions with special spin quantum numbers can be constructed directly
based of the solutions in $2+1$ dimensions. For this, we must choose the $z$%
-component of the polarization pseudovector $S^{3}$ as the spin operator in $%
3+1$ dimensions. This is a new result not only for the magnetic-solenoid
field background, but also for the pure AB field. The choice of $S^{3}$ as
the spin operator is convenient from different standpoints. For example, the
solutions with arbitrary momentum $p^{3}$ are the eigenvectors of the
operator $S^{3}$. This allows explicitly separating the spin and coordinate
variables in $3+1$ dimensions and reducing the problem to the problem of the
self-adjoint extension for the radial Hamiltonian only.\ Moreover, the
boundary conditions in such a representation do not violate the translation
invariance along the natural direction which is the magnetic-solenoid field
direction. Using von Neumann's theory of the self-adjoint extensions of
symmetric operators, we have constructed the \thinspace\ self-adjoint
extensions of the Dirac Hamiltonian with the magnetic-solenoid field and
obtained a one-parameter family and a two-parameter family of admissible
self-adjoint boundary conditions in respective $2+1$ dimensions and $3+1$
dimensions. The complete orthonormal sets of solutions thus have been found.
We have determined the energy spectra dependent on the extension parameter $%
\Theta $ for different \thinspace\ self-adjoint extensions. In addition, we
have for the first time described the solutions of the Dirac equation with
the regularized magnetic-solenoid field in detail. We have considered an
arbitrary magnetic field distribution inside the finite-radius solenoid and
shown that the extension parameters $\Theta =-\mathrm{sgn}(\Phi )\pi /2$ in $%
2+1$ dimensions and $\Theta _{+1}=\Theta _{-1}=-\mathrm{sgn}(\Phi )\pi /2$
in $3+1$ dimensions correspond to the limiting case $R\rightarrow 0$ of the
regularized magnetic-solenoid field. The finite radius solenoid
regularization was also considered for the spinless particle case. It was
demonstrated that in contrast to the spinning particle case, the
corresponding (as $R\rightarrow 0$) radial functions are regular for the
arbitrary magnetic field inside the solenoid.

\section{Acknowledgments}

The authors are grateful to I. Tyutin for useful discussions. D.M.G. thanks
CNPq and FAPESP for permanent support. A.A.S. and S.P.G thank FAPESP for
support. B.L.V. thanks LSS-1578.2003.2 and RFBR 02-02-16946 for support.

\setcounter{section}{0} \renewcommand{\thesection}{\Alph{section}}

\section{Appendix}

1.The Laguerre function $I_{n,m}(x)$ is defined by 
\begin{equation}
I_{n,m}(x)=\sqrt{\frac{\Gamma \left( 1+n\right) }{\Gamma \left( 1+m\right) }}%
\frac{\exp \left( -x/2\right) }{\Gamma \left( 1+n-m\right) }x^{\left(
n-m\right) /2}\Phi (-m,n-m+1;x)\,.  \label{4.1}
\end{equation}
Here, $\Phi \left( a,b;x\right) $ is the confluent hypergeometric function
in a standard definition (see \cite{GR94}, 9.210). Let $m$ be a non-negative
integer number; then the Laguerre function is related to the Laguerre
polynomials $L_{m}^{\alpha }(x)$ (\cite{GR94}, 8.970, 8.972.1) by 
\begin{eqnarray}
&&I_{m+\alpha ,m}(x)=\sqrt{\frac{m!}{\Gamma \left( m+\alpha +1\right) }}%
e^{-x/2}x^{\alpha /2}L_{m}^{\alpha }(x)\,,  \label{4.42} \\
&&L_{m}^{\alpha }(x)=\frac{1}{m!}e^{x}x^{-\alpha }\frac{d^{m}}{dx^{m}}%
e^{-x}x^{m+\alpha }\,.  \label{4.5}
\end{eqnarray}
Using the well-known properties of the confluent hypergeometric function ( 
\cite{GR94}, 9.212; 9.213; 9.216), we can easily obtain the following
relations for the Laguerre functions: 
\begin{eqnarray}
&&2\sqrt{x(n+1)}I_{n+1,m}(x)=(n-m+x)I_{n,m}(x)-2xI_{n,m}^{\prime }(x)\,,
\label{4.11} \\
&&2\sqrt{x(m+1)}I_{n,m+1}(x)=(n-m-x)I_{n,m}(x)+2xI_{n,m}^{\prime }(x)\,,
\label{4.12} \\
&&2\sqrt{xn}I_{n-1,m}(x)=(n-m+x)I_{n,m}(x)+2xI_{n,m}^{\prime }(x)\,,
\label{4.13} \\
&&2\sqrt{xm}I_{n,m-1}(x)=(n-m-x)I_{n,m}(x)-2xI_{n,m}^{\prime }(x)\,.
\label{4.16}
\end{eqnarray}
Using the properties of the confluent hypergeometric function, we obtain the
representation 
\begin{equation}
I_{n,m}(x)=\sqrt{\frac{{\Gamma (}1+n{)}}{{\Gamma (}1+m{)}}}{\frac{{\exp }%
\left( x/2\right) }{{\Gamma (}1+n-m{)}}}x^{\frac{{n-m}}{2}}\Phi
(1+n,1+n-m;-x)  \label{4.24}
\end{equation}
and the relation (\cite{GR94}, 9.214)

\begin{equation}
I_{n,m}(x)=(-1)^{n-m}I_{m,n}(x),\;n-m\;\mathrm{integer}\,.  \label{4.25}
\end{equation}
The functions $I_{\alpha +m,m}(x)$ satisfy the orthonormality condition, 
\begin{equation}
\int_{0}^{\infty }I_{\alpha +n,n}\left( x\right) I_{\alpha +m,m}\left(
x\right) dx=\delta _{m,n}\,,  \label{4.41}
\end{equation}
which follows from the corresponding properties of the Laguerre polynomials
( \cite{GR94}, 7.414.3). The set of the Laguerre functions 
\begin{equation*}
I_{\alpha +m,m}(x),\;m=0,1,2...\,\,,\;\alpha >-1
\end{equation*}
is complete in the space of square integrable functions on the half-line ($%
x\geq 0$), 
\begin{equation}
\sum_{m=0}^{\infty }I_{\alpha +m,m}(x)I_{\alpha +m,m}(y)=\delta \left(
x-y\right) \,.  \label{ap18}
\end{equation}

2. The function $\psi _{\lambda ,\alpha }(x)$ is even with respect to the
index $\alpha $, 
\begin{equation}
\psi _{\lambda ,\alpha }\left( x\right) =\psi _{\lambda ,-\alpha }\left(
x\right) \,.  \label{4.138}
\end{equation}
It can be expressed via the confluent hypergeometric functions, 
\begin{eqnarray}
&&\psi _{\lambda ,\alpha }\left( x\right) =e^{-\frac{x}{2}}\left[ \frac{%
\Gamma \left( -\alpha \right) x^{\frac{\alpha }{2}}}{\Gamma \left( \frac{%
1-\alpha }{2}-\lambda \right) }\Phi \left( \frac{1+\alpha }{2}-\lambda
,1+\alpha ;x\right) \right.  \notag \\
&&\left. +\frac{\Gamma \left( \alpha \right) x^{-\frac{\alpha }{2}}}{\Gamma
\left( \frac{1+\alpha }{2}-\lambda \right) }\Phi \left( \frac{1-\alpha }{2}%
-\lambda ,1-\alpha ;x\right) \right] \,,  \label{4.139}
\end{eqnarray}
or, using (\ref{4.1}), via the Laguerre functions, 
\begin{eqnarray}
&&\psi _{\lambda ,\alpha }\left( x\right) =\frac{\sqrt{\Gamma \left(
1+n\right) \Gamma \left( 1+m\right) }}{\sin \left( n-m\right) \pi }\left(
\sin n\pi I_{n,m}\left( x\right) -\sin m\pi I_{m,n}\left( x\right) \right)
\,,  \notag \\
&&\alpha =n-m,\;2\lambda =1+n+m,\;n=\lambda -\frac{1-\alpha }{2},\;m=\lambda
-\frac{1+\alpha }{2}\,.  \label{4.140}
\end{eqnarray}
The relations 
\begin{eqnarray}
&&\psi _{\lambda ,\alpha }\left( x\right) =\sqrt{x}\psi _{\lambda -\frac{1}{2%
},\alpha -1}\left( x\right) +\frac{1+\alpha -2\lambda }{2}\psi _{\lambda
-1,\alpha }\left( x\right) \,,  \notag \\
&&\psi _{\lambda ,\alpha }\left( x\right) =\sqrt{x}\psi _{\lambda -\frac{1}{2%
},\alpha +1}\left( x\right) +\frac{1-\alpha -2\lambda }{2}\psi _{\lambda
-1,\alpha }\left( x\right) \,,  \notag \\
&&2x\psi _{\lambda ,\alpha }^{\prime }\left( x\right) =\left( 2\lambda
-1-x\right) \psi _{\lambda ,\alpha }\left( x\right) +\frac{1}{2}\left(
2\lambda -1-\alpha \right) \left( 2\lambda -1+\alpha \right) \psi _{\lambda
-1,\alpha }\left( x\right) \,,  \notag \\
&&2x\psi _{\lambda ,\alpha }^{\prime }\left( x\right) =\left( \alpha
-x\right) \psi _{\lambda ,\alpha }\left( x\right) +\left( 2\lambda -1-\alpha
\right) \sqrt{x}\psi _{\lambda -\frac{1}{2},\alpha +1}\left( x\right)  \notag
\\
&&\,=\left( x-2\lambda -1\right) \psi _{\lambda ,\alpha }-2\psi _{\lambda
+1,\alpha }\,  \label{4.144}
\end{eqnarray}
are valid for the functions $\psi _{\lambda ,\alpha }\left( x\right) $.\ The
direct consequence of these relations is 
\begin{eqnarray}
&&A_{\alpha }\psi _{\lambda ,\alpha }\left( x\right) =\frac{2\lambda
-1+\alpha }{2}\psi _{\lambda -\frac{1}{2},\alpha -1}\left( x\right)
,\;A_{\alpha }^{+}\psi _{\lambda -\frac{1}{2},\alpha -1}\left( x\right)
=\psi _{\lambda ,\alpha }\left( x\right) \,,  \notag \\
&&A_{\alpha }=\frac{x+\alpha }{2\sqrt{x}}+\sqrt{x}\frac{d}{dx}\,,\;A_{\alpha
}^{+}=\frac{x+\alpha -1}{2\sqrt{x}}-\sqrt{x}\frac{d}{dx}\,.  \label{4.145}
\end{eqnarray}
Using the well-known asymptotic behavior of the Whittaker function (\cite
{GR94}, 9.227), we have 
\begin{equation}
\psi _{\lambda ,\alpha }\left( x\right) \sim x^{\lambda -\frac{1}{2}}e^{-%
\frac{x}{2}}\,,\;x\rightarrow \infty ;\;\;\psi _{\lambda ,\alpha }\left(
x\right) \sim \frac{\Gamma \left( |\alpha |\right) }{\Gamma \left( \frac{%
1+|\alpha |}{2}-\lambda \right) }x^{-\frac{|\alpha |}{2}}\,,\;\alpha \neq
0\,,\;x\sim 0\,.  \label{4.147}
\end{equation}
The function $\psi _{\lambda ,\alpha }\left( x\right) $ is correctly defined
and infinitely differentiable for $0<x<\infty $ and for any complex $\lambda
,\alpha .$ In this respect , we note that the Laguerre function is not
defined for negative integer $n$ and $m.$ In the particular cases where one
of the numbers $n$ or $m$ is non-negative and integer, the function $\psi
_{\lambda ,\alpha }\left( x\right) $ coincides (up to a constant factor)
with the Laguerre function.

According to (\ref{4.147}), the functions $\psi _{\lambda ,\alpha }\left(
x\right) $ are square integrable on the interval $0\leq x<\infty $ whenever $%
|\alpha |<1$. This is not true for $|\alpha |\geq 1.$ The corresponding
integrals for $\alpha \neq 0$ are calculated (see 7.611 in (\cite{GR94} )): 
\begin{eqnarray}
&&\int\limits_{0}^{\infty }\psi _{\lambda ,\alpha }\left( x\right) \psi
_{\lambda ^{\prime },\alpha }\left( x\right) \,dx=\frac{\pi }{\left( \lambda
^{\prime }-\lambda \right) \sin \alpha \pi }\left\{ \left[ \Gamma \left( 
\frac{1+\alpha -2\lambda ^{\prime }}{2}\right) \Gamma \left( \frac{1-\alpha
-2\lambda }{2}\right) \right] ^{-1}\right.  \notag \\
&&\left. -\left[ \Gamma \left( \frac{1-\alpha -2\lambda ^{\prime }}{2}%
\right) \Gamma \left( \frac{1+\alpha -2\lambda }{2}\right) \right]
^{-1}\right\} \,,\quad |\alpha |<1\,,  \label{4.148} \\
&&\int\limits_{0}^{\infty }|\psi _{\lambda ,\alpha }\left( x\right) |^{2}dx=%
\frac{\pi }{\sin \alpha \pi }\frac{\psi \left( \frac{1+\alpha -2\lambda }{2}%
\right) -\psi \left( \frac{1-\alpha -2\lambda }{2}\right) }{\Gamma \left( 
\frac{1+\alpha -2\lambda }{2}\right) \Gamma \left( \frac{1-\alpha -2\lambda 
}{2}\right) },\quad |\alpha |<1\,.  \label{4.150}
\end{eqnarray}
Here, $\psi (x)$ is the logarithmic derivative of the $\Gamma $-function ( 
\cite{GR94}, 8.360). In the general case, the functions $\psi _{\lambda
,\alpha }\left( x\right) \;$and $\psi _{\lambda ^{\prime },\alpha }(x),$ $%
\lambda ^{\prime }\neq \lambda ,$ are not orthogonal, as it follows from (%
\ref{4.148}).

\section{Appendix}

We here present modifications of some foregoing formulas for the case $B<0$.

1. The spectrum $\omega $ corresponding to the functions $\phi _{m,l,\sigma
}(r)$ is 
\begin{equation}
\omega =\left\{ 
\begin{array}{l}
2\gamma \left( m-l+1-\mu \right) ,\;{}l-\left( 1+\sigma \right) /2<0 \\ 
2\gamma \left( m+\left( 1-\sigma \right) /2\right) ,\;{}l-\left( 1+\sigma
\right) /2\geq 0
\end{array}
\right. ,  \label{abe23}
\end{equation}
and the spectrum $\omega $ corresponding to the functions $\phi _{m,\sigma
}^{ir}(r)$ is 
\begin{equation}
\omega =\left\{ 
\begin{array}{l}
2\gamma \left( m+1-\mu \right) ,\;\sigma =-1 \\ 
2\gamma m,\;\sigma =1
\end{array}
\right. .  \label{abe24}
\end{equation}
These expressions are the modifications of Eqs. (\ref{abe21}) and (\ref
{abe22}) for the case $B<0$.

2.We consider the spinors $u_{l}\left( r\right) $ satisfying (\ref{abe26}).\
In the case $\omega =0$, these are 
\begin{equation}
u_{0,l}(r)=\left( 
\begin{array}{l}
\phi _{0,l,1}(r) \\ 
0
\end{array}
\right) \,,\mathrm{\;}{}l\geq 1;\;\;u_{0}^{II}(r)=\left( 
\begin{array}{l}
\phi _{0,1}^{ir}(r) \\ 
0
\end{array}
\right) ,\;{}l=0\,.  \label{abe29}
\end{equation}
In the case $\omega \neq 0$, these are 
\begin{eqnarray}
&&u_{m,l,\pm }(r)=\left( 
\begin{array}{l}
\phi _{m,l,1}(r) \\ 
\mp i\phi _{m,l,-1}(r)
\end{array}
\right) ,\mathrm{\;}{}l\leq -1\;,{}\;\omega =2\gamma \left( m-l+1-\mu
\right) \,,  \notag \\
&&u_{m+1,l,\pm }(r)=\left( 
\begin{array}{l}
\phi _{m+1,l,1}(r) \\ 
\pm i\phi _{m,l,-1}(r)
\end{array}
\right) ,\mathrm{\;}{}l\geq 1\;,{}\;\omega =2\gamma \left( m+1\right) \,, 
\notag \\
&&u_{m+1,\pm }^{II}(r)=\left( 
\begin{array}{l}
\phi _{m+1,1}^{ir}(r) \\ 
\pm i\phi _{m,0,-1}(r)
\end{array}
\right) ,\mathrm{\;}{}l=0\;,{}\;\omega =2\gamma \left( m+1\right) \,,  \notag
\\
&&u_{m,\pm }^{I}(r)=\left( 
\begin{array}{l}
\phi _{m,0,1}(r) \\ 
\mp i\phi _{m,-1}^{ir}(r)
\end{array}
\right) ,\mathrm{\;}{}l=0\;,{}\;\omega =2\gamma \left( m+1-\mu \right) \,.
\label{abe30}
\end{eqnarray}
The presented expressions are the modifications of Eqs. (\ref{abe27}) and (%
\ref{abe28}) for the case $B<0$.

3. In the case $\omega =0$, the only positive energy solutions (particles)
of Eq. (\ref{abe12}) are possible. These solutions coincide with the
corresponding spinors $u$ up to a normalization constant: 
\begin{equation}
_{+}\psi _{0,l}(r)=Nu_{0,l}(r){},\;l\geq 1\,;\;\;_{+}\psi
_{0}^{II}(r)=Nu_{0}^{II}(r){},\;l=0\,.  \label{abe33}
\end{equation}
Thus, the particles have the rest energy level, whereas the energy spectrum
of antiparticles begins from $_{-}\varepsilon =-\sqrt{M^{2}+2\gamma }$.

4. For the case $B<0$, the relations for the irregular spinors $u_{\omega
,\sigma }(r)$ similar to the relations (\ref{abe35}) are 
\begin{equation}
\Pi u_{\omega ,-1}(r)=i\sqrt{2\gamma }u_{\omega ,1}(r),\;\;\Pi u_{\omega
,1}(r)=-i\frac{\omega }{\sqrt{2\gamma }}u_{\omega ,-1}(r)\,.  \label{abe36}
\end{equation}

\end{document}